\documentclass[journal]{IEEEtran}

\usepackage{cite}
\usepackage{amsthm}
\usepackage{bm}
\usepackage{amsmath,amssymb,amsfonts}
\usepackage{mathtools}
\usepackage{algorithm,algorithmic}
\usepackage{textcomp}
\usepackage[dvipsnames]{xcolor}
\usepackage{booktabs}
\def\BibTeX{{\rm B\kern-.05em{\sc i\kern-.025em b}\kern-.08em
    T\kern-.1667em\lower.7ex\hbox{E}\kern-.125emX}}

\newcommand{\ba}{\begin{array}}
\newcommand{\ea}{\end{array}}
\newcommand{\be}{\begin{displaymath}}
\newcommand{\ee}{\end{displaymath}}
\newcommand{\ben}{\begin{equation}}
\newcommand{\een}{\end{equation}}
\newcommand{\bena}{\begin{eqnarray}}
\newcommand{\eena}{\end{eqnarray}}
\newcommand{\beqa}{\begin{eqnarray*}}
\newcommand{\enqa}{\end{eqnarray*}}

\newcommand{\bc}{\begin{center}}
\newcommand{\ec}{\end{center}}
\newcommand{\bi}{\begin{itemize}}
\newcommand{\ei}{\end{itemize}}
\newcommand{\benu}{\begin{enumerate}}
\newcommand{\eenu}{\end{enumerate}}
\newcommand{\bdes}{\begin{description}}
\newcommand{\edes}{\end{description}}
\newcommand{\bt}{\begin{tabular}}
\newcommand{\et}{\end{tabular}}

\newcommand \Phibf{\boldsymbol{\Phi}}

\newcommand \phibf{\boldsymbol{\phi}}

\newcommand \bbf{{\bf b}}

\newcommand \fbf{{\bf f}}
\newcommand \gbf{{\bf g}}
\newcommand \hbf{{\bf h}}

\newcommand \pbf{{\bf p}}

\newcommand \xbf{{\bf x}}

\newcommand \zbf{{\bf z}}

\newcommand \Abf{{\bf A}}

\newcommand \Fbf{{\bf F}}
\newcommand \Gbf{{\bf G}}
\newcommand \Hbf{{\bf H}}
\newcommand \Ibf{{\bf I}}

\newcommand \Zbf{{\bf Z}}

\newcommand{\circlambda}{\mbox{$\Lambda$
             \kern-.85em\raise1.5ex
             \hbox{$\scriptstyle{\circ}$}}\,}

\newcommand{\tr}{\mathop{\rm tr}}

\makeatletter
\newcommand{\ALOOP}[1]{\ALC@it\algorithmicloop\ #1%
	\begin{ALC@loop}}
	\newcommand{\ENDALOOP}{\end{ALC@loop}\ALC@it\algorithmicendloop}

\makeatother



\usepackage{geometry}
\ifCLASSINFOpdf
\else
\fi
\usepackage{scalerel}
\usepackage{tikz}
\usetikzlibrary{svg.path}

\definecolor{orcidlogocol}{HTML}{A6CE39}
\tikzset{
  orcidlogo/.pic={
    \fill[orcidlogocol] svg{M256,128c0,70.7-57.3,128-128,128C57.3,256,0,198.7,0,128C0,57.3,57.3,0,128,0C198.7,0,256,57.3,256,128z};
    \fill[white] svg{M86.3,186.2H70.9V79.1h15.4v48.4V186.2z}
                 svg{M108.9,79.1h41.6c39.6,0,57,28.3,57,53.6c0,27.5-21.5,53.6-56.8,53.6h-41.8V79.1z M124.3,172.4h24.5c34.9,0,42.9-26.5,42.9-39.7c0-21.5-13.7-39.7-43.7-39.7h-23.7V172.4z}
                 svg{M88.7,56.8c0,5.5-4.5,10.1-10.1,10.1c-5.6,0-10.1-4.6-10.1-10.1c0-5.6,4.5-10.1,10.1-10.1C84.2,46.7,88.7,51.3,88.7,56.8z};
  }
}

\newcommand\orcidicon[1]{\href{https://orcid.org/#1}{\mbox{\scalerel*{
\begin{tikzpicture}[yscale=-1,transform shape]
\pic{orcidlogo};
\end{tikzpicture}
}{|}}}}

\usepackage{hyperref}

\begin{document}

\title{Jammer Mitigation in Absorptive RIS-Assisted Uplink NOMA}

\author{Azadeh Tabeshnezhad\textsuperscript{\orcidicon{0000-0003-0140-0076}},~\IEEEmembership{Member,~IEEE}, 
Yuqing Zhu\textsuperscript{\orcidicon{0000-0002-1442-1063}},~\IEEEmembership{Member,~IEEE},
Artem Vilenskiy\textsuperscript{\orcidicon{0000-0002-7701-799X}},~\IEEEmembership{Senior Member,~IEEE},
Ly V. Nguyen\textsuperscript{\orcidicon{0000-0002-2682-4118}},~\IEEEmembership{Senior Member,~IEEE},
A. Lee Swindlehurst\textsuperscript{\orcidicon{0000-0002-0521-3107}},~\IEEEmembership{Fellow,~IEEE},
Tommy Svensson\textsuperscript{\orcidicon{0000-0002-2579-9002}},~\IEEEmembership{Senior Member,~IEEE}
\thanks{Azadeh Tabeshnezhad, Yuqing Zhu, Artem Vilenskiy, and Tommy Svensson are with the Department of Electrical Engineering, Chalmers University of Technology, 412 96 Gothenburg, Sweden (e-mail:azadeh.tabeshnezhad@chalmers.se; yuqingz@chalmers.se; artem.vilenskiy@chalmers.se; tommy.svensson@chalmers.se).}
\thanks{Van Ly Nguyen is with the Department of Electrical Engineering and Computer Science at the University of Kansas (KU) 66045, USA (e-mail:vanly.nguyen@ku.edu).}
\thanks{Arnold Lee Swindlehurst is with the Department of Electrical Engineering and Computer Science at the University of California at Irvine (UCI) 92697, USA (e-mail:swindle@uci.edu).}
}

\maketitle

\begin{abstract}
Non-orthogonal multiple access (NOMA) is a promising technology for next-generation wireless communication systems due to its enhanced spectral efficiency. However, wireless communication is facing increasing requirements for security. To that end, jamming mitigation using multi-antennas has emerged as an important research topic. 
In this paper, we consider an uplink NOMA system with a reconfigurable intelligent surface (RIS) that assists the uplink users and, at the same time, mitigates the jammer. Our goal is to minimize the total users’ transmitted power under signal-to-interference-plus-noise ratio constraints at the base station. To be effective, typically a high-dimensional RIS is needed, leading to a large optimization problem, which in general faces convergence problems.
We propose an iterative algorithm for this high-dimensional non-convex optimization problem that converges with a jammer comprising as many as 64 antennas, and an RIS with 128 elements. More specifically, we introduce a design and optimize the performance of an absorptive RIS (A-RIS). Compared to a standard RIS, we show that an A-RIS can dramatically reduce the users’ required transmit power and successfully mitigate the jammer. The A-RIS is in particular useful in cases when the number of jammer antennas is of the same order as the number of A-RIS elements.
\end{abstract}


\IEEEpeerreviewmaketitle

\section{Introduction}
\IEEEPARstart Non-orthogonal multiple access (NOMA) is a promising technique for next-generation radio access. In contrast to conventional orthogonal multiple access (OMA) schemes, where each user is served using a single orthogonal resource block, NOMA is of particular interest because it allows multiple users to share the same orthogonal time, frequency, spatial, and code-domain resource blocks. NOMA has significant advantages over traditional OMA schemes, with higher effective bandwidth, support for massive connectivity, and reduced transmission latency \cite{b1}.   

Generally, NOMA is divided into two main classes: power-domain and code-domain \cite{b2}. Power-domain NOMA exploits situations where the users have different path loss levels. The idea behind power-domain NOMA is that the users nearer the base station (BS) can employ successive interference cancellation (SIC) to remove the strong signals destined for the remote users before decoding their own signal \cite{b3}.

\begin{figure}
    \centering
    \includegraphics[height=3.5cm]{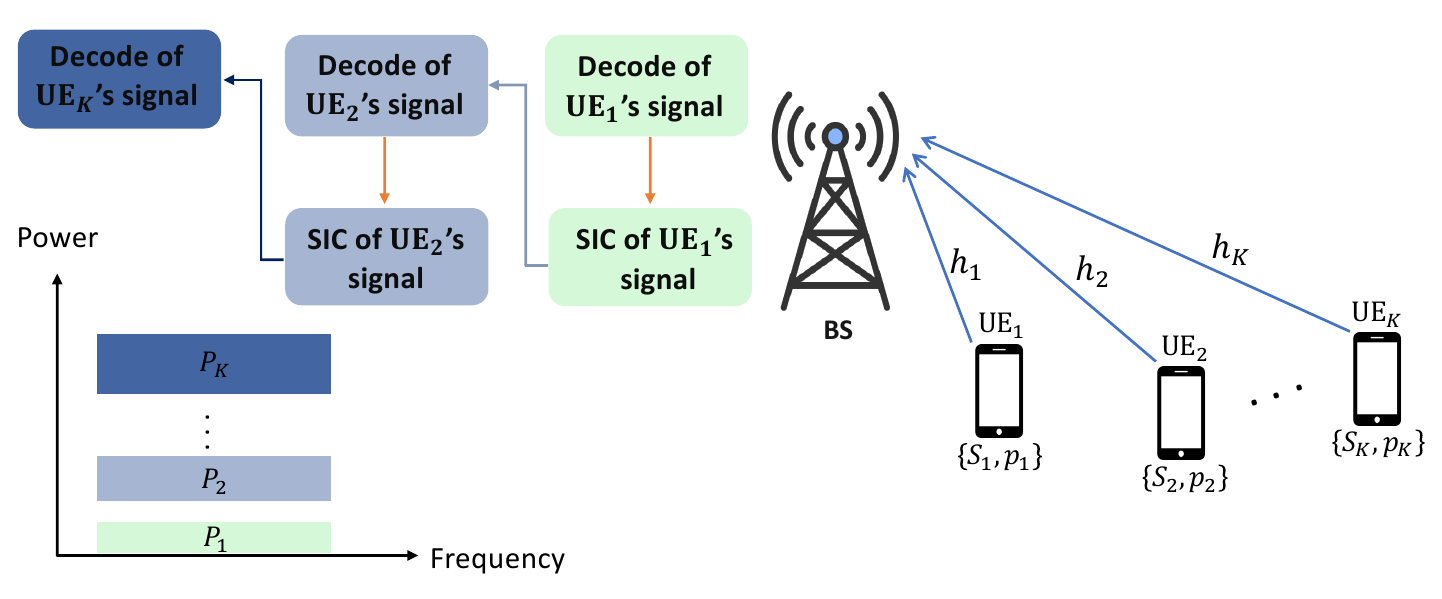}
    \caption{Illustration of uplink power-domain NOMA with $K$ users sending messages $S_1$, $S_2$, $\dots$, and $S_K$ with power 
    $p_1$, $p_2$, $\dots$, and $p_K$ respectively.}
    \label{fig1}
\end{figure}

In uplink NOMA, which is the focus of this paper, the BS can similarly employ SIC to remove stronger user signals before decoding a given signal of interest. However, as illustrated in Fig.~\ref{fig1}, the SIC condition needs to include both the individual users' transmit powers, as well as the channel gain of the individual users' channels sharing the same NOMA resource. Thus, to effectively implement SIC, the transmit power levels of the uplink users are important. There is still substantially less work on uplink NOMA compared to the downlink case, and there is also a lack of work addressing the reliability of NOMA in the presence of active jamming. For either downlink or uplink NOMA, when the users and the BS only have one antenna, dealing with jammers is a challenge since neither the users (for the downlink) nor the BS (for the uplink) may have sufficient spatial degrees of freedom (DoFs) to cancel the jamming \cite{b4}. In such cases, the jamming can severely degrade the NOMA performance even when decoding is performed by or for the strongest user.

Recently, the use of RIS has emerged as a unique technology for improving both the spectral and energy efficiency of wireless networks. An RIS consists of an array of elements whose reflective properties can be individually controlled, enabling some degree of control of the wireless propagation environment. In wireless communications, RIS can be easily deployed on various outdoor and indoor structures, including walls, vehicle windows, advertising billboards, etc., due to their use of small, low-cost, and lightweight elements \cite{b5}. RIS can provide additional channel paths to construct stronger combined channels with significant differences in strength \cite{b6}.

In contrast to active amplify-and-forward relays that usually operate in half-duplex mode, the RIS functions in full-duplex mode and only reflects the received signals as a passive array, without the expenditure of any radio-frequency energy, which enables spectrum-efficient and cost-effective communications \cite{b7}, \cite{b8}, \cite{b9}. An RIS can increase the strength of received signals through beamforming gain and constructive interference, eliminate interfering signals through destructive interference or null-steering, and provide virtual line-of-sight paths to overcome blockages. For single-antenna NOMA systems in the presence of a jammer, an RIS has the potential to provide a large number of spatial DoFs that could allow the network to cancel the jamming signal, essentially creating a null channel from the jammer to either the single-antenna BS or single-antenna users. This is the key RIS capability that will be studied in this paper in the context of jamming mitigation in uplink NOMA.

In a conventional RIS implementation, it is the phase shift of the reflection coefficient of each element that is adjusted in order to achieve the desired effect on the wireless channels.
More recently, researchers have also studied RIS architectures where both the phase shift and the attenuation or absorption of the reflection coefficient of each element can be controlled individually, referred to as absorptive RIS (A-RIS) \cite{b10}.
In this latter case, the energy absorbed by the A-RIS can be refracted to directions on the other side of the surface \cite{b11}, \cite{b12} sampled using an active RF receiver for channel estimation or sensing  \cite{b13}, \cite{b14},\cite{b15} or simply dissipated.
In \cite{b16}, the interference mitigation problem using A-RIS in two spectrum sharing scenarios is investigated: under spectral coexistence of radar and communication systems, and spectrum sharing in device-to-device (D2D) communications. The results show that A-RIS significantly outperforms non-absorptive RIS for interference suppression scenarios. Similarly, the interference cancellation ability of A-RIS was exploited to support joint D2D and cellular communications in \cite{b17}.

RIS technology has been applied in many different types of wireless communications scenarios, including NOMA. However, while the use of conventional phase-shift-only RIS has been proposed for NOMA applications with the goal of improving spectral efficiency, to the best of our knowledge, there is no prior work reported on using A-RIS with NOMA, nor on using RIS to mitigate the impact of external interference (e.g., jamming) on NOMA performance. NOMA is a vulnerable technology due to the use of shared resources since the decoding performance depends on sufficient differences in the users' received power levels, which in the uplink involves adapting the transmit power levels to the channel conditions. However, jamming can reduce the efficiency of NOMA due to the severe loss of signal-to-noise ratio (SNR) \cite{b18}.

\subsection{Related Works}
Several recent works have studied the use of RIS in NOMA applications. For example, an RIS-assisted NOMA system was investigated in \cite{b19} to achieve a tradeoff between sum rate and power consumption. 
In \cite{b20}, joint power allocation, hybrid beamforming, and phase shift optimization for downlink multi-user RIS-assisted mm-wave NOMA was designed to maximize the sum rate using alternating manifold optimization (AMO) and successive convex approximation (SCA) methods. 
In \cite{b21}, the maximum sum rate achieved by optimizing the RIS phase shifts subject to the user power constraints was studied for NOMA. In \cite{b22}, NOMA was implemented in a multi-cell scenario assisted by multiple RISs to minimize the transmit power in the uplink. 
An RIS-aided multi-unmanned aerial vehicle (UAV) system with NOMA was studied in \cite{b23}, with the goal of minimizing the system's power consumption.
In \cite{b24}, RIS-aided multi-cell NOMA was shown to improve the coverage probability using stochastic geometry methods and two different channel models. Furthermore, many other research problems in the context of RIS-assisted NOMA have been studied using different optimization methods in \cite{b25}, \cite{b26}.

Relatively little work has been done on optimizing NOMA performance in the presence of a jammer. In \cite{b27}, optimal user grouping for NOMA was proposed to overcome the impact of jamming and improve the sum rate. In \cite{b28}, a mobile access point or a UAV was exploited together with joint power control to mitigate the effect of a jamming attack and increase the reliability of the communication. Furthermore, anti-jamming precoding was proposed in \cite{b29} to minimize the total transmit power in an uplink MIMO-NOMA system. In \cite{b30}, transmit beamforming together with the use of an RIS and artificial noise was proposed to enhance the secrecy of a NOMA system. Several works have studied the use of RIS for anti-jamming in \cite{JMA+Tang2021},\cite{RAR+Sun2022}, and \cite{SNS+Ji2023}.
However, except for our initial work in \cite{b25}, none of the previous works cited above have considered the use of an RIS for mitigating jamming for an uplink NOMA system.

\subsection{Motivations and Contributions}
As already mentioned above, the focus of this paper is on the use of RIS, and in particular A-RIS, to mitigate the influence of a jammer on an uplink NOMA system. Our objective is to minimize the sum transmit power of the uplink NOMA users under quality of service (QoS) constraints in the presence of a jammer. This problem formulation is well aligned with 6G key requirements on sustainability, trustworthiness, and digital inclusion \cite{URB+21}. The main contributions of our work are summarized as follows:
\begin{itemize}
    \item We show that an uplink power-domain NOMA system assisted by an A-RIS can play a key role in mitigating interference from a sophisticated multi-antenna jammer. Compared to an uplink NOMA system without A-RIS assistance, we show a gain of more than 28 dB in the required sum transmit power by the uplink users. These gains come from the combined capability of an A-RIS to not only mitigate the jammer interference but also enable substantial beamforming and interference coordination gains for the uplink NOMA users. Unlike conventional RIS, we show that the nature of A-RIS introduces additional new important degrees of freedom in optimizing the system performance while keeping power consumption low.
    \item We propose a novel joint optimization problem formulation that minimizes the total transmit power of user terminals under constraints on the A-RIS response and the individual signal-to-interference-plus-noise ratio (SINR) at the BS for each of the users. This problem has not been studied in the literature before, and it turns out to be a highly complex non-convex optimization problem due to the strong coupling between user power control and the A-RIS phase-shift and gain design.
    \item To efficiently solve this problem, we develop an iterative algorithm that combines Linear Programming (LP) to find the user powers, ensuring optimal power allocation in each iteration and sequential convex relaxation via Dinkelbach’s algorithm \cite{Dinkelbach} to efficiently optimize the A-RIS response, leveraging fractional programming principles. The proposed algorithm does not rely on conventional convex approximations such as SDR (Semidefinite Relaxation), making it more computationally scalable and adaptable to large-scale A-RIS, which is a key enabler for realistic A-RIS deployments. Our approach effectively balances power minimization and interference mitigation.
    \item We present the results of several simulations that demonstrate the convergence of the algorithm and the effectiveness of the A-RIS in mitigating the jammer and controlling the multi-user interference, enabling the system to operate with relatively low sum transmit power. These results show that the proposed algorithm converges very well and can handle A-RIS with a large number of elements and a jammer with many antenna elements. This is very important since the passive nature of A-RIS implies that more elements are typically needed compared to conventional relays or APs.
    
\end{itemize}

\textit{Organization}: The rest of the paper includes the following sections. In Section~\ref{sec:SystemModel}, we detail the targeted scenario and the A-RIS-assisted NOMA system model, and we define the resulting optimization problem. In Section~\ref{sec:Absorb}, we provide details on the implementation of the A-RIS, which motivates our system model and simulation assumptions. Section~\ref{sec:Dinkelbach Algorithm} then provides details for our proposed optimization algorithm that alternates between LP and the Dinkelbach algorithm combined with semi-definite relaxation (SDR) to minimize the total power transmitted from the users to the BS. Section~\ref{sec:PerformanceAnalysis} then presents numerical results that validate the performance of the algorithm in challenging scenarios, followed by a discussion on the relevance of the results for a corresponding downlink scenario. Finally, we summarize our findings in Section~\ref{sec:Conclusion}. 

\textit{Notation}: Scalars, vectors, and matrices are respectively written with italic, bold lowercase, and bold uppercase characters (e.g., $h$, $\hbf$, $\Hbf$). The conjugate and transpose of vectors and matrices are respectively expressed by the superscripts $*$, $\top$ (e.g., $\zbf^*$, $\Zbf^*$, $\zbf^\top$, $\Zbf^\top$). In addition, diag$\mathrm{(\fbf)}$ denotes the diagonal matrix whose diagonal elements are taken from the vector $\mathrm{\fbf}$, $\|.\|$ returns the norm of a vector, and $|.|$ provides the magnitude of a complex number. The operator $\mathrm{tr}(\Zbf)$ yields the trace of a matrix, and $\mathbb{C}$ denotes the space of complex numbers.

\section{System Model and Problem Formulation}\label{sec:SystemModel}

In this section, we first describe the system and signal model for our considered A-RIS-aided NOMA system and provide assumptions regarding the channel state information (CSI) available to the BS and the jammer. 

\subsection{System Model}\label{AA}

\textbf{\begin{figure}
    \centering
    \includegraphics[height =6cm]{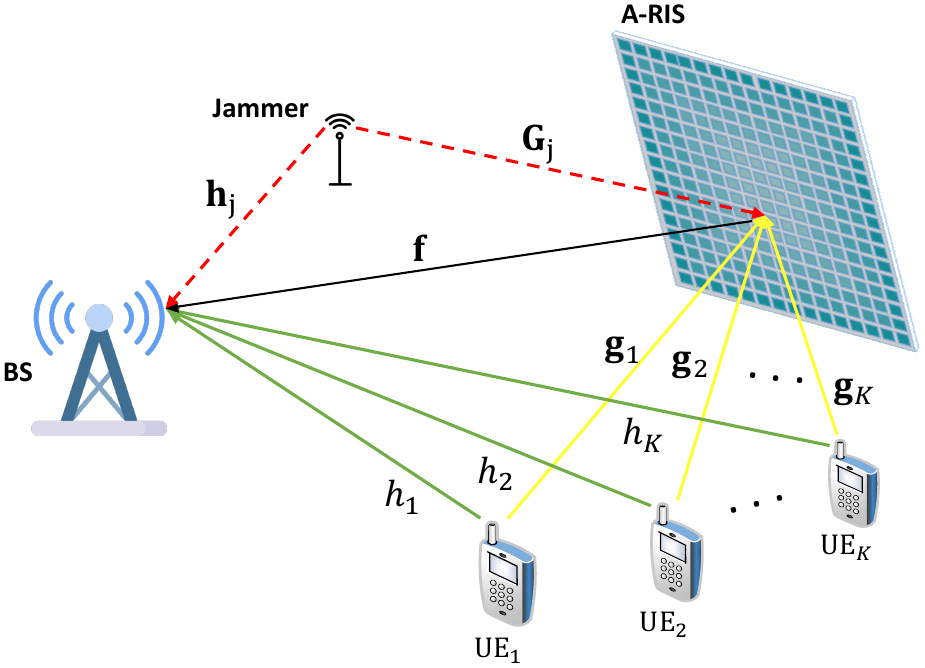}
    \caption{Illustration of an A-RIS-aided uplink NOMA system with $K$ users and a jammer.}
    \label{fig2}
\end{figure}
}

We consider an A-RIS-aided uplink NOMA system with a BS equipped with a single-antenna, an A-RIS with $N$ elements, a jammer with $M$ antennas, and $K$ single-antenna users employing power-domain NOMA, as shown in Fig.~\ref{fig2}. For this scenario, the received signal at the BS can be written as
\begin{equation} \label{eu_1}
r = \sum_{i=1}^K \left( h_i + \fbf^\top \Phibf\gbf_i \right) x_i + \left( \hbf_\text{j}^\top + \fbf^\top \Phibf\Gbf_\text{j}\right) \xbf_\text{j} +n_\text{r},
\end{equation}
where $h_i \in \mathbb{C}^{1\times 1}$ denotes the direct channels between the users and the BS, $x_i$ is the symbol transmitted by user $i$, $\fbf \in \mathbb{C}^{N\times 1}$ denotes the channel vector between the A-RIS and the BS, $\bm {g}_i\in \mathbb{C}^{N\times 1}$ is the channel vector between user $i$ and the A-RIS, $\bm {h}_\text{j} \in \mathbb{C}^{M\times 1}$ represents the channel vector between the jammer and the BS,  $\Gbf_\text{j} \in \mathbb{C}^{N\times M}$ denotes the channel matrix between the jammer and A-RIS, $\xbf_\text{j} \in \mathbb{C}^{M\times 1}$ is the signal transmitted by the jammer, and $n_\text{r}$ $\sim$ $\mathcal{CN}$ $(0,\sigma^2)$ is additive white Gaussian noise (AWGN). The matrix $\Phibf = \text{diag} \{\beta_1 e^{j\theta_1}, \beta_2 e^{j\theta_2},...,\beta_N e^{j\theta_N}\}\in \mathbb{C}^{N\times N}$ is a diagonal matrix consisting of the adjustable reflection coefficients of the A-RIS, where for the $n$-th element of an A-RIS we have an adjustable phase $\theta_n\in [0,2\pi]$ and an adjustable amplitude $\beta_n \le \beta_\text{max} \le 1$.  We define the transmit power of each user as $p_i = E(|x_i|^2)$. We assume that the BS has the CSI for all the channels in~\eqref{eu_1}.

Assume the $K$ users are indexed in the order of declining channel gains, and $p_1\vert h_1 \vert^2 \geq  p_2\vert h_2 \vert^2  \geq ... \geq p_k\vert h_k \vert^2$, consistent with the situation illustrated in Fig.~\ref{fig1}. The BS then first decodes the signal for UE$_1$ treating the multi-user interference of the signal from the other users as noise, and then subtracts it from the received signal $r$ when decoding the signal from UE$_2$. Thus, the SINR for decoding UE$_1$ is given by
\begin{equation} \label{eu_2}
    \gamma_1 = \frac{p_1\vert h_1+\fbf^\top\Phibf\gbf_1\vert^2}{\smashoperator\sum_{i=2}^{K} p_{i}\vert h_{i}+\fbf^\top\Phibf\gbf_{i}\vert^2+\sigma_\text{j}^2+\sigma^2} ,  
\end{equation}
where we let $\sigma_\text{j}^2$ denote the power of the jammer signal received by the BS. This term will be derived in Section~\ref{sec:Jammer} below. Under the assumption of perfect SIC of UE$_1$'s signal, i.e., without error propagation, the SINR for UE$_2$ can be written as
\begin{equation} \label{eu_3}
   \gamma_2 = \frac{p_2 \vert h_2+\fbf^\top\Phibf\gbf_2\vert^2}{\smashoperator\sum_{i=3}^{K} p_{i}\vert h_{i}+\fbf^\top\Phibf\gbf_{i}\vert^2+\sigma_\text{j}^2+\sigma^2}  \; .
\end{equation}
Continuing to assume a perfect SIC for UE$_2$ and each of the subsequent user's signals with no error propagation, the SINR for the last user, UE$_K$ can be written as 
\begin{equation} \label{eu_4}
   \gamma_K = \frac{p_K\vert  h_K +\fbf^\top \Phibf \gbf_K \vert^2 }{\sigma_\text{j}^2 +\sigma^2} \; .
\end{equation}
The goal to be addressed is the design of the A-RIS response $\Phibf$ to minimize the sum of the users' transmit powers $\sum_i p_i$ while satisfying constraints on the SINRs $\gamma_i, i=1, \cdots, K.$

\subsection{Jammer Modeling}
\label{sec:Jammer}
We assume an intelligent jammer that is aware of $\Phibf$ and the CSI of all the channels affected by the jammer, i.e. \{$\hbf_\text{j}$, $\fbf$, $\Gbf_\text{j}$\}. The objective of the jammer is to choose $\xbf_\text{j}$ such that the interference power received at the BS is maximized. 
In other words, the jammer designs $\xbf_\text{j}$ based on the following criterion:
\begin{align} \label{eu_5}
\sigma_\text{j}^2 = &\max_{\xbf_\text{j}} \; \| \left(\hbf_\text{j}^\top + \fbf^\top\Phibf\Gbf_\text{j}\right) \xbf_\text{j} \|^2 \\ &\mbox{\rm s.t.} \quad E \left(\|\xbf_\text{j}\|^2\right) \le P_\text{j} \nonumber \; ,
\end{align}
where $P_\text{j}$ is the jammer's maximum transmit power. The solution to this problem is simply $\xbf_\text{j}=(\sqrt P_\text{j}/\rho) \left(\hbf_\text{j}^\top+\fbf^\top\Phibf\Gbf_\text{j}\right)^H$, where $\rho=\|\hbf_\text{j}^\top+\mathbf{f}^\top\Phibf\Gbf_\text{j}\|$. Thus, the term due to jamming in the denominator of the SINR expressions in (\ref{eu_2})-(\ref{eu_4}) becomes:
\begin{equation} \label{eu_6}
 \sigma_\text{j}^2 = P_\text{j}\| \hbf_\text{j}^\top+\fbf^\top\Phibf\Gbf_\text{j} \|^2 \; .
\end{equation}
Note that the jammer transmit power is fixed at $P_\text{j}$, and thus the power transmitted per jammer antenna decreases as the number of jammer antennas $M$ increases. However,  the amount of jamming power that reaches the BS will in general increase with $M$ due to the coherent beamforming gain, thus the need for the intervention of the A-RIS. We also note that our subsequent analysis and algorithm development are relevant to the case where the jammer does not possess CSI. For example, suppose the jammer simply broadcasts spatially white Gaussian noise, i.e., $\xbf_\text{j} \sim {\cal{CN}}(0,\frac{P_\text{j}}{M}\Ibf)$. In this case, the power received at the BS becomes $\sigma_\text{j}^2 = (P_\text{j}/M) \| \hbf_\text{j}^\top+\fbf^\top\Phibf\Gbf_\text{j} \|^2$, or simply $1/M$ the power in the case of coherent jamming. Thus, results obtained for an intelligent jammer with a given $P_\text{j}$ will be the same as those for a ``dumb'' jammer with $M$ times more power.

\section{Absorptive RIS} \label{sec:Absorb}
As discussed above, recent work has considered RIS implementations in which not all energy is reflected from the surface. In prior work, the non-reflected energy is either transmitted or ``refracted'' to the other side of the RIS, or it is demodulated and sampled for channel estimation or sensing purposes. Here we simply assume that the RIS absorbs an adjustable fraction of the incoming energy at each element, without assuming that the absorbed energy is used for any other purpose. 

In Section~\ref{sec:PerformanceAnalysis}, we show that the ability to adjust the magnitude of the reflected power at each RIS element provides additional DoFs that are particularly useful in situations like the one we consider in this work, in which interference mitigation is required. 

\subsection{Discussion on RIS Implementation With Both Amplitude and Phase Control} \label{sec:ARIScontrol}

As already introduced in Section~\ref{sec:SystemModel}, the A-RIS model assumes that the reflection coefficient of each RIS element can be described as $\beta_n e^{j\theta_n}$, where $0 \le \beta_n \le 1$ describes the amplitude of the reflected signal component, whereas, in the conventional RIS, it is assumed that $\beta_n =1 $.
In practice, depending on the update rate of the A-RIS, due to the limited capacity of the BS to A-RIS control channel, the values of $\{\beta_n \}$, and $\{ \theta_n \}$, may have to be quantized. However, in this work, we assume the possibility of continuously controlling both and ignoring the impact of quantization. Such control information with a high sampling rate and quantization accuracy is feasible in less dynamic scenarios. The reason is twofold; the constraint $0 \le \beta_n \le 1$ has numerical advantages as it is convex, and as we show below, it is possible to independently adjust phase and amplitude in a practical design of the A-RIS elements.

\subsection{Practical Design of A-RIS} \label{sec:ARISdesign}
Compared to widely studied phase-controlled RIS designs, A-RIS designs with concurrent amplitude and phase reconfigurability are less prevalent due to the challenges in physical implementation and availability of tunable components \cite{Y. Saifullah}. Nevertheless, in recent years, the higher flexibility and versatility provided by combined amplitude-phase control has sparked growing research interest in this technology in the antenna, optics, and materials communities \cite{J. Liao}. As an example, take the reconfiguration mechanism based on using semiconductor \emph{p-i-n} and varactor diodes. A \emph{p-i-n} diode can not only function as a microwave switch to generate a discrete phase shift but can also operate as a variable resistor by applying a forward current to bias the diode's junction to the transition state. In this case, the junction resistance can vary by several orders of magnitude \cite{J. Liao}. When an RIS unit cell (UC) combines a \emph{p-i-n} diode with a varactor diode, exploiting the control of the reverse-biased junction capacitance, a variable amplitude-phase RIS reflection response can be engineered. Recently, such a reconfigurable RIS UC was experimentally demonstrated, allowing independent and continuous amplitude-phase control \cite{Y. Zhu1}.

\begin{figure}[!t]
\centerline{\includegraphics[height =5cm]{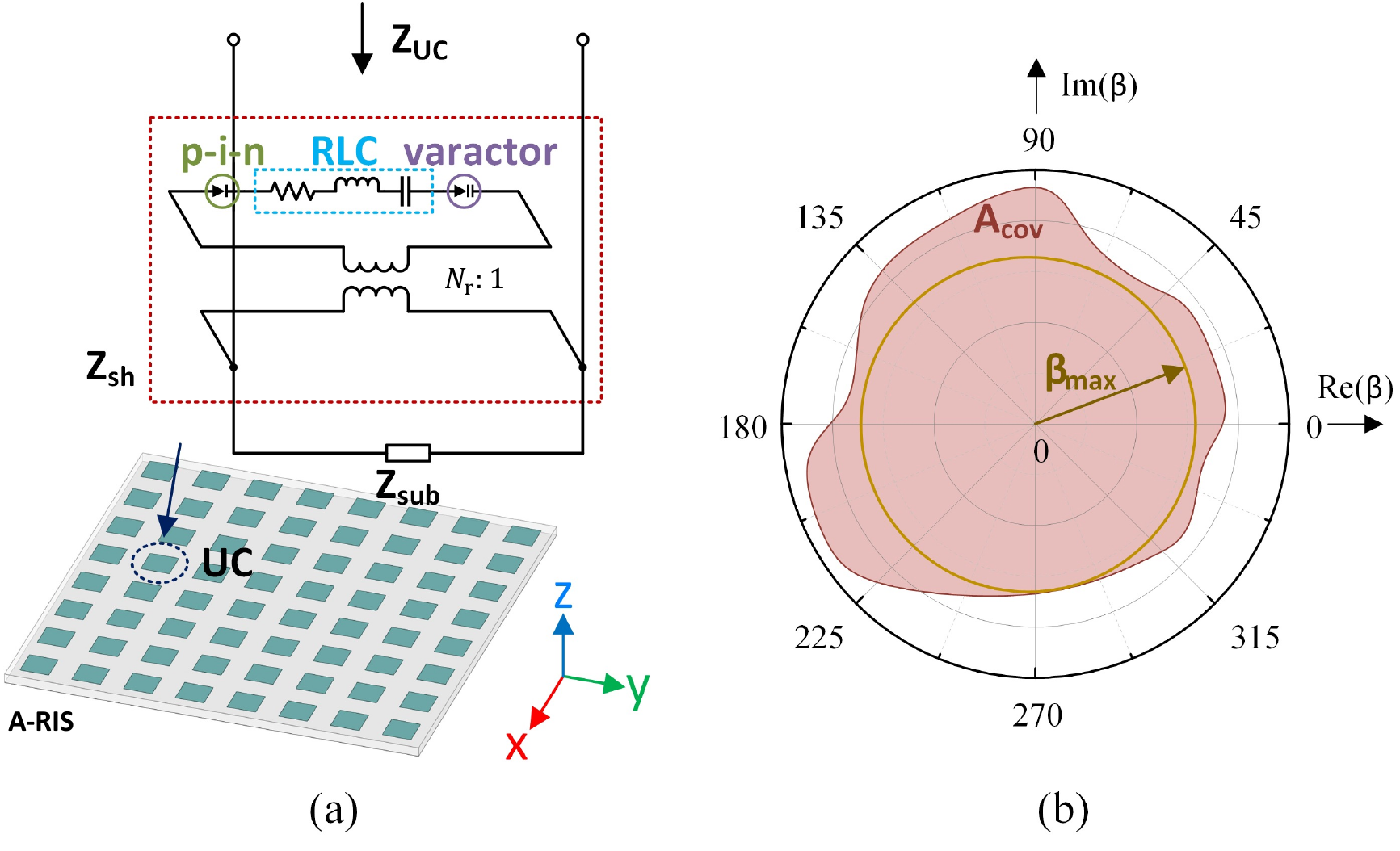}}
\caption{A-RIS with both amplitude and phase control. (a) Equivalent circuit model of the UC. (b) UC reflection coefficient coverage in the complex plane.}
\label{fig_RIS_Implementation}
\end{figure}

In most cases, an RIS UC is implemented as a scattering metal topology distributed on a grounded dielectric substrate and carrying controllable components (i.e., diodes). The UC typically has polarization axes, e.g., $x$-axis in the local RIS coordinate system, defining the $E$-field orientation for which the UC can control reflection. Omitting technical details, a conceptual circuit diagram of a single-polarized, amplitude-and-phase controllable RIS UC is illustrated in Fig.~\ref{fig_RIS_Implementation}(a). For a plane wave illuminating the RIS from the free-space side, the UC can be seen as a single-port network with some complex input impedance $Z_\mathrm{UC}$, also known as the opaque surface impedance \cite{G. Minatti}. The latter can be decomposed into an equivalent sheet impedance (controllable by diode biasing) on the air-dielectric interface $Z_\mathrm{sh}$ and a substrate input impedance $Z_\mathrm{sub}$, connected in parallel, i.e., $Z_\mathrm{UC} = Z_\mathrm{sh} || Z_\mathrm{sub}$. The sheet impedance is modeled as a circuit containing \emph{RLC} elements, representing passive metal parts of the topology, \emph{p-i-n} and varactor diodes, which are connected to the input UC nodes through an impedance transformer with a turns ratio $\sqrt{N_\mathrm{r}}$ Fig.~\ref{fig_RIS_Implementation}(a). A value for $N_\mathrm{r}$ is defined by a sheet current distribution. At the same time, $Z_\mathrm{sub}$ can be regarded as the impedance of a short-circuited transmission line segment. 
This way, the UC reflection property can be described by a single scalar reflection coefficient $\beta = (Z_\mathrm{UC} - Z_\mathrm{s})/(Z_\mathrm{UC} + Z_\mathrm{s})$, where $Z_\mathrm{s}$ is the wave impedance of the incident plane wave.

\subsection{Maximum Reflection versus Phase of A-RIS}
By continuously applying a forward bias current to the \emph{p-i-n} diode and a reverse bias voltage to the varactor diode, values for $\beta$ for all operating states can be mapped onto the complex plane, forming an effective UC coverage region, having area $A_\mathrm{cov}$ Fig.~\ref{fig_RIS_Implementation}(b). The coverage region represents the achievable UC $\beta$ range, thus characterizing its reflection control capability over both amplitude and phase. In particular, a circle with a maximum radius $\beta_\mathrm{max}$ fully contained inside the coverage region can be introduced for quantitative analysis. Within this circular region, the A-RIS provides a full phase range of (0--360)$^{\circ}$ and an amplitude range of $(0-\beta_\mathrm{max})$.
Note that the $\beta_\mathrm{max}$ is typically less than 1, due to inherent losses in the UC. These losses include ohmic losses of a copper UC topology, substrate dielectric losses, and additional losses caused by the junction resistance and parasitic parameters of the diodes employed.

\section{Optimization Method} \label{sec:Dinkelbach Algorithm}
Based on the UC design outlined in the previous section, here we address optimization of the A-RIS response in order to mitigate the impact of a jammer in an uplink NOMA system. We will assume A-RIS elements that are independently tunable within the disk of radius $\beta_\mathrm{max}$ depicted in Fig.~\ref{fig_RIS_Implementation}(b). Clearly, there are additional $\beta$ values greater than $\beta_\mathrm{max}$ that can be achieved for certain phase values $\theta$, but we ignore this possibility in the discussion below to maintain a convex set of parameters.

\subsection{Problem Formulation for \texorpdfstring{$K$} {TEXT} NOMA Uplink Users}
There are several ways to illustrate the benefit of the A-RIS in the problem under consideration. Here we study the problem of minimizing the total transmit power of the users such that given SINR QoS constraints of the users are met in the presence of a jammer. The optimization problem for $K$ users can be written as follows:
\begin{subequations}\label{eu_General}
\begin{gather}\label{eu_7a}
\min_{\{p_k\},\Phibf}~\sum_{k=1}^K p_k\\ \label{eu_7b}
\text{s.t.}~\mathbf{C1}:p_k\geq0,~ k=1,\,\ldots,\, K\\ \label{eu_7c}
\mathbf{C2}:\frac{p_k\vert h_k+\fbf^\top\Phibf\gbf_k\vert^2}{\smashoperator\sum_{i=k+1}^ {K} p_{i}\vert h_{i}+\fbf^\top\Phibf\gbf_{i}\vert^2+\sigma_\text{j}^2+\sigma^2}\geq T_k, \\ k=1,\,\ldots,\,K  \nonumber\\
\mathbf{C3}:\vert\Phibf(n,n)\vert\leq \beta_\mathrm{max},~n=1,\,\ldots,\,N. \label{eu_7d}
\end{gather}
\end{subequations} 
Constraint {\bf C1} ensures that each of the transmit powers is non-negative, constraint {\bf C2} corresponds to the desired individual SINR constraint for each user UE$_k$, $k=1,\cdots, K$, and constraint {\bf C3} indicates that the configuration of each A-RIS element $\{\beta_1 e^{j\theta_1},\beta_2 e^{j\theta_2},\,\ldots,\,\beta_N e^{j\theta_N}\}$ should lie within a circle of maximum radius $\beta_\mathrm{max}$ in the complex plane.

Note that in our optimization problem formulation above, we ensure robust SIC by incorporating SINR constraints into the optimization problem, which enforce a minimum threshold to guarantee the required QoS for each user. By setting these constraints appropriately, the system automatically achieves the conditions necessary for successful SIC. Therefore, even without explicitly defining SIC constraints, our approach inherently ensures their fulfillment.

We see from~(\ref{eu_General}) that the problem of interest is non-linear, non-convex, and high-dimensional, and cannot be solved directly. Therefore, we take the steps outlined below to transform it into a more tractable form that is amenable to optimization. We begin by noting that the weakest user must satisfy constraint {\bf C2} with equality, otherwise $p_K$ could be further reduced to achieve a lower value of the criterion function. Thus, we have 
\begin{equation} \label{eu_8}
p_K\vert h_K+\fbf^\top\Phibf\gbf_K\vert^2 = T_K(P_\text{j}\Vert\hbf_\text{j}^\top+\fbf^\top\Phibf\Gbf_\text{j}\Vert^2+\sigma^2).
\end{equation}
Substituting (\ref{eu_8}) into (\ref{eu_7a}) and after some simple modifications, the original optimization problem in (\ref{eu_General}) becomes
\begin{subequations} \label{eu_9G}
\begin{gather} \label{eu_9a}
\min_{p_1,p_2,\,\ldots,\,p_{K-1},\Phibf}~p_1+\ldots+\frac{T_K(P_\text{j}\Vert\hbf_\text{j}^\top+\fbf^\top\Phibf\Gbf_\text{j}\Vert^2+\sigma^2)}{\vert h_K+\fbf^\top\Phibf\gbf_K\vert^2} \\  \label{eu_9b}
\text{s.t.}~ p_k\geq0, ~ k=1,\,\ldots,\, K\\ 
p_k\vert h_k+\fbf^\top\Phibf\gbf_k\vert^2 \geq T_k (P_\text{j}\Vert\hbf_\text{j}^\top+\fbf^\top\Phibf\Gbf_\text{j}\Vert^2+ \notag \\ \label{eu_9c} \sum_{i=k+1}^Kp_{i}\vert h_{i}+\fbf^\top\Phibf\gbf_{i}\vert^2+\sigma^2), ~ k=1,\,\ldots,\, K-1 \\   
\vert\Phibf(n,n)\vert\leq\beta_\mathrm{max},~n=1,\,\ldots,\,N. \label{eu_9d}
\end{gather}
\end{subequations}

Next, we convert the squared magnitude terms in the SINR constraints to quadratic forms involving the diagonal elements $\phibf = [ \beta_1 e^{j\theta_1} \beta_2 e^{j\theta_2} \ldots \beta_N e^{j\theta_N}]^T$ of $\Phibf$:
\begin{subequations} \label{eu_10}
\begin{gather}\label{eu_10a}
\vert h_k+\fbf^\top\Phibf\gbf_k\vert^2=\bar\phibf^H\Fbf_k\bar\phibf,\\ \label{eu_10b}
\Vert\hbf_\text{j}^\top+\fbf^\top\Phibf\Gbf_\text{j}\Vert^2=\bar\phibf^H\Fbf_\text{j}\bar\phibf\; ,   
\end{gather}
\end{subequations}
where $\bar\phibf = [\phibf^\top \; 1]^T$,
\begin{align} \label{eu_11}
    &\Fbf_k = \begin{bmatrix}
        \zbf_k^*\zbf_k^\top & \zbf_k^* h_k \\
        h_k^* \zbf_k^\top & |h_k|^2
    \end{bmatrix},
        \; \Fbf_{\text{j}} = \begin{bmatrix}
        \Zbf^*\Zbf^\top & \Zbf^* \hbf_{\text{j}} 
         \\ 
        \hbf_{\text{j}}^H\Zbf^\top & \|\hbf_{\text{j}}\|^2
    \end{bmatrix},
\end{align}
$\zbf_k = (\fbf \odot \gbf_k)$, and $\Zbf = \operatorname{diag}(\fbf)\Gbf_{\text{j}}$. Note that
$\Fbf_k \in \mathbb{C}^{N+1\times N+1}$, $\Fbf_\text{j} \in \mathbb{C}^{N+1\times N+1}$,
$\zbf_i\in \mathbb{C}^{N\times 1}$ and $\Zbf\in \mathbb{C}^{N\times 1}$.
With the above definitions, the optimization problem in (\ref{eu_9G}) can be rewritten as
\begin{subequations}\label{eu_12}
\begin{gather}
\min_{p_1,p_2,..,p_{K-1},\bar\phibf}~p_1+...+\frac{T_K(P_\text{j}\bar\phibf^H\Fbf_\text{j}\bar\phibf+\sigma^2)}{\bar\phibf^H\Fbf_K\bar\phibf}\\
\text{s.t.}~p_k\geq0, ~ k=1,\,\ldots,\, K\\ \nonumber
p_k \bar\phibf^H \Fbf_k\bar\phibf\geq T_k (P_\text{j}\bar\phibf^H \Fbf_\text{j}\bar\phibf+\sum_{i=k+1}^Kp_{i} \bar\phibf^H \Fbf_i\bar\phibf+\sigma^2 ), \\ ~ k=1,\,\ldots,\, K-1 \\ 
\vert\bar\phibf(n)\vert\leq\beta_\mathrm{max},~n=1,\cdots,N \\
 \vert\bar\phibf(N+1)\vert=1.
\end{gather}
\end{subequations}

\subsection{Proposed Solution}
We see from~\eqref{eu_12} that if the user powers $\{p_k\}$ are known, the optimization requires solving a fractional quadratically constrained quadratic programming (QCQP) problem. Thus, to solve~\eqref{eu_12}, we propose an algorithm that iterates between (1) the use of LP to find the users' transmit powers, and (2) the Dinkelbach algorithm together with SDR \cite{SDR} to solve for the optimal A-RIS response. We first introduce a new variable, $\bar\Phibf=\bar\phibf\bar\phibf^H$, which is a rank-one Hermitian semidefinite matrix. Using the SDR approach, we convert the quadratic constraints into linear matrix inequalities and we relax the rank-one constraint as follows:
\begin{subequations}\label{eu_13}
\begin{gather} \label{eu_13a}
\min_{p_1,p_2,..,p_{K-1},\bar\Phibf}~p_1+...+\frac{T_K(P_\text{j}\tr (\Fbf_\text{j} \bar\Phibf)+\sigma^2)}{\tr(\Fbf_K \bar\Phibf)}\\ \label{eu_13b}
\text{s.t.}~ p_k\geq0, ~ k=1,\,\ldots,\, K\\ \nonumber
p_k\tr (\Fbf_k \bar \Phibf) \geq T_k (P_\text{j}\tr (\Fbf_\text{j} \bar \Phibf)+\sum_{i=k+1}^Kp_{i}\tr (\Fbf_i\bar \Phibf)+\sigma^2), \nonumber \\
~ k=1,\,\ldots,\, K-1 \\ 
\label{eu_13d}
\vert\bar\Phibf(n,n)\vert\leq\beta_\mathrm{max},~ n=1,\cdots,N \\
 \vert\bar\Phibf(N+1,N+1)\vert=1. \label{eu_13e}
\end{gather}
\end{subequations}

With $\bar\Phibf$ fixed, the $\{ p_k \}$ can be found directly using LP:
\begin{subequations}\label{eu_14}
\begin{gather}
\min \mathbf{1}^\top\pbf\\
\text{s.t.}~\Abf\pbf \leq \bbf\\
\pbf \geq \mathbf{0}
\end{gather}
\end{subequations}
where $\pbf = [p_1,\ldots,p_K]^T$, and 
        \begin{subequations}  \label{eu_15} \setlength\arraycolsep{-0.01cm}
        \begin{align}
        \Abf &= \begin{bmatrix} \label{eu_15a}
        -\tr(\Fbf_1\bar\Phibf) &~T_1\tr(\Fbf_2\bar\Phibf) & ~~T_1\tr(\Fbf_3\bar\Phibf) & ... & T_1\tr(\Fbf_K\bar\Phibf)  \\
        0 &~ -\tr(\Fbf_2\bar\Phibf) &~~T_2\tr(\Fbf_3\bar\Phibf) & ... & T_2\tr(\Fbf_K\bar\Phibf) \\
        0 & 0 &~~ -\tr(\Fbf_3\bar\Phibf) & ... & T_3\tr(\Fbf_K\bar\Phibf)\\
        \vdots & \vdots & \vdots & \ddots & \vdots \\
         0 & 0 & 0 & \cdots & -\tr(\Fbf_K \bar\Phibf)
    \end{bmatrix} \\ \label{eu_15b} 
    \bbf& = \begin{bmatrix}
     -T_1(P_\text{j} \tr(\Fbf_1 \bar\Phibf) - \sigma^2) \\
    -T_2(P_\text{j}\tr(\Fbf_2 \bar\Phibf) - \sigma^2) \\
    -T_3(P_\text{j}\tr(\Fbf_3 \bar\Phibf) - \sigma^2)  \\
    \vdots \\
    -T_K(P_\text{j} \tr(\Fbf_K \bar\Phibf) - \sigma^2) 
    \end{bmatrix}.  
    \end{align}
    \end{subequations} 

Next, \eqref{eu_13} is solved assuming $\pbf$ is fixed. Eliminating terms from the objective function that do not depend on $\bar{\Phibf}$, we have
\begin{subequations}\label{eu_16}
\begin{gather} \label{eu_16a}
\min_{\bar\Phibf}~\frac{T_K(P_\text{j}\tr (\Fbf_\text{j} \bar\Phibf)+\sigma^2)}{\tr(\Fbf_K \bar\Phibf)}\\ 
\text{s.t.}~ p_k\tr (\Fbf_k \bar \Phibf) \geq T_k(P_\text{j}\tr (\Fbf_\text{j} \bar \Phibf)+\notag \\  \sum_{i=k+1}^Kp_{i}\tr (\Fbf_i\bar \Phibf)+\sigma^2),
~ k=1,\,\ldots,\, K-1 \\
\vert\bar\Phibf(n,n)\vert\leq\beta_\mathrm{max},~ n=1,\cdots,N \\\vert\bar\Phibf(N+1,N+1)\vert=1. 
\end{gather}
\end{subequations}
Thanks to the relaxation, what remains is a fractional programming problem with linear inequality constraints, so we turn to the Dinkelbach algorithm to optimize over $\bar\Phibf$. To do so, we introduce a slack variable $\lambda$ and rewrite problem~(\ref{eu_16}) as
\begin{subequations}\label{eu_17}
\begin{gather} \label{eu_17a}
\min_{\lambda,\bar\Phibf}~{T_K(P_\text{j}\tr(\Fbf_\text{j}\bar\Phibf)+\sigma^2)}-\lambda {\tr(\Fbf_K\bar\Phibf)}\\ 
\text{s.t.}~
p_k\tr (\Fbf_k \bar \Phibf) \geq T_k (P_\text{j}\tr (\Fbf_\text{j} \bar \Phibf)+ \notag\\ 
\sum_{i=k+1}^Kp_{i}\tr (\Fbf_i\bar \Phibf)+\sigma^2),
~ k=1,\,\ldots,\, K-1 \\ 
\vert\bar\Phibf(n,n)\vert\leq\beta_\mathrm{max},~ n=1,\cdots,N
\\\vert\bar\Phibf(N+1,N+1)\vert=1.
\end{gather}
\end{subequations}
The Dinkelbach procedure operates by iteratively solving for $\lambda$ and $\bar{\Phibf}$. For fixed $\bar{\Phibf}$, the solution for $\lambda$ can be obtained in closed form:

\begin{equation}\label{eu_18}
\lambda=\frac{T_K(P_\text{j}\tr(\Fbf_\text{j}\bar\Phibf)+\sigma^2)}{\tr(\Fbf_K\bar\Phibf)}.   
\end{equation}
Once $\lambda$ is found, the optimization for $\bar{\Phibf}$ is convex. This inner iteration between $\lambda$ and $\bar{\Phibf}$ continues until a convergence criterion is met. As a final step, the rank of the solution must be examined as a result of the SDR. Let $\bar\Phibf^\star$ denote the solution after convergence. If the rank of $\bar\Phibf^\star$ is $1$, the optimal solution for $\bar\phibf$ is obtained directly using an eigenvalue decomposition. Otherwise, a rank-one approximation or Gaussian randomization must be employed to find $\bar{\phibf}^*$. The optimal A-RIS response $\phibf^\star$ is found via the normalization
$\phibf^\star = \bar\phibf^\star(1:N)/\bar{\phibf}^*(N+1)$.

\subsection{Algorithm Initialization}
Our empirical results indicate that the procedure outlined above critically depends on proper initialization. We have found that the following initialization procedure works well, based on the observation that, for sufficiently large $N$, the optimal $\bar{\Phibf}$ tends to eliminate the contribution of the jammer at the BS:
\begin{enumerate}
\item Solve the following convex problem to find a $\bar{\Phibf}^{(-1)}$ that minimizes the jammer power at the BS:
\begin{subequations}\label{init1}
\begin{gather} 
\bar{\Phibf}^{(-1)} = \arg\min_{\bar\Phibf}~\tr(\Fbf_\text{j}\bar\Phibf) \\ 
\text{s.t.}~
\vert\bar\Phibf(n,n)\vert\leq\beta_\mathrm{max},~ n=1,\cdots,N
\\\vert\bar\Phibf(N+1,N+1)\vert=1.
\end{gather}
\end{subequations}
The solution to~\eqref{init1} is typically not unique.
\item Use $\bar{\Phibf}^{(-1)}$ in~\eqref{eu_14} to compute the initial set of user powers $\mathbf{p}^{0}$.
\item Find the initial A-RIS response $\bar{\Phibf}^{(0)}$ as the one that minimizes the jammer power at the BS, {\em and} satisfies the SINR constraints based on $\mathbf{p}^{(0)}$:
\begin{subequations}\label{init2}
\begin{gather}
\bar{\Phibf}^{(0)} = \arg\min_{\bar\Phibf}~\tr(\Fbf_\text{j}\bar\Phibf) \\ 
\text{s.t.}~
p_k^{(0)}\tr (\Fbf_k \bar \Phibf) \geq T_k (P_\text{j}\tr (\Fbf_\text{j} \bar \Phibf)+ \notag\\ 
\sum_{i=k+1}^Kp_{i}^{(0)}\tr (\Fbf_i\bar \Phibf)+\sigma^2),
~ k=1,\,\ldots,\, K-1 \\ 
\vert\bar\Phibf(n,n)\vert\leq\beta_\mathrm{max},~ n=1,\cdots,N
\\\vert\bar\Phibf(N+1,N+1)\vert=1.
\end{gather}
\end{subequations}
This is also a convex problem and easily solvable.
\end{enumerate}
With the initialization procedure defined above, the overall algorithm is summarized as Algorithm~1 below.

\section{Performance Analysis} \label{sec:PerformanceAnalysis}
In this section, we analyze the optimized performance of the A-RIS-assisted uplink NOMA system in Fig.~\ref{fig2} using Algorithm~\ref{alg:PLS} in various scenarios.

\subsection{Performance for Various Number of Users with Fixed A-RIS Location} \label{sec:FixedARISLocation}

\begin{algorithm}[tpb]
\caption{LP + Dinkelbach Approach}
\label{alg:PLS}
    \begin{algorithmic}[1]
        \REQUIRE Channel state information, $P_j$, $\sigma^2$, thresholds $\{T_k\}$, $\beta_\mathrm{max}$, convergence thresholds $\epsilon_1, \epsilon_2$
        \ENSURE Optimal A-RIS configuration $\phibf^\star$, minimum user transmit powers $\{p_k^\star\}$
        \STATE \textbf{Initialization} 
        \STATE Solve (\ref{init1}) to find $\bar{\Phibf}^{(-1)}$ 
        \STATE Compute $\mathbf{p}^{0}$ using $\bar{\Phibf}^{(-1)}$ in~\eqref{eu_14} 
        \STATE Compute $\bar{\Phibf}^{(0)}$ using $\mathbf{p}^{0}$ in (\ref{init2})
        \STATE $q=1$
        \REPEAT 
        \STATE Compute $\mathbf{p}^{(q)}$ using $\bar{\Phibf}^{(q-1)}$ in~\eqref{eu_14}
        \STATE $\mathrm{diff1}(q)=\mathbf{p}^{(q)}-\mathbf{p}^{(q-1)}$
        \STATE Initialize $\lambda^{(0)}$ using $\bar{\Phibf}^{(q-1)}$ in~\eqref{eu_18}
        \STATE $\ell=1$
        \REPEAT 
        \STATE Compute $\bar{\Phibf}^{(q,\ell)}$ using $\pbf^{(q)}$ and $\lambda^{(\ell-1)}$ in~\eqref{eu_17}
        \STATE Compute $\lambda^{(\ell)}$ using $\bar{\Phibf}^{(q,\ell)}$ in~\eqref{eu_18}
        \STATE $\mathrm{diff2}(\ell)=\lambda^{(\ell)}-\lambda^{(\ell-1)}$
        \STATE $\ell \longleftarrow \ell+1$
        \UNTIL $\mathrm{diff2}(\ell-1) < \epsilon_2$
        \STATE $\bar{\Phibf}^{(q)}=\bar{\Phibf}^{(q,\ell-1)}$
        \STATE $q \longleftarrow q+1$
        \UNTIL $\mathrm{diff1}(q-1) < \epsilon_1$
        \STATE Set $\mathbf{p}^\star=\mathbf{p}^{(q-1)}$
        \STATE Set $\bar\Phibf^\star$ = $\bar\Phibf^{(q)}$
        \STATE Perform a rank-one decomposition or Gaussian randomization on $\bar{\Phibf}^\star$ to find $\bar\phibf^*$, then set $\phibf^\star = \bar\phibf^\star(1:N)/\bar{\phibf}^*(N+1)$.
        \STATE \textbf{return}  
        \end{algorithmic}
\end{algorithm}

Here we study the performance of the proposed algorithm in terms of the sum total uplink transmit power for $K=2$, $K=3$ \textcolor{blue}{and $K=4$} users, for various numbers of jammer antennas and A-RIS elements, assuming a fixed location of the A-RIS. We also assess the benefit and characterize the optimized behavior of the absorption capability of the A-RIS in terms of the average reflection coefficients of the A-RIS elements.

In Table~\ref{table1} we show our simulation parameters. Here we assume a fixed network, whereas in Section~\ref{sec:VariousARISLocations} we relax this by studying the performance for various A-RIS locations. We focus on a scenario with static rather than random channel gains to ensure that near-far relationship between the users remains the same throughout the simulation trials, which (together with the SINR constraints in the optimization) enables us to fix the decoding order rather than adding this as an additional complication to the simulation. The benefit of the A-RIS in mitigating the jamming for the NOMA system is ultimately independent of this assumption, so the purpose is mainly to simplify our experiments.  Note that for numerical stability in the optimization, we normalize the noise at the BS as $\sigma^2=1$ and scale the path losses correspondingly. The chosen path loss values are representative for a network with inter-node distances of a few tens of meters, assuming path loss exponent $\alpha=3$.

\begin{table}
    \caption{Summary of Simulation Parameters. 
    \label{table1}}
    \centering
    \begin{tabular}{l|l|l}
    \bf{Variable}    &  \bf{Description}            & \bf{Value} \\ \hline \hline
    $P_\text{j}$     &  Normalized jammer antenna power (Watt) & 10   \\ \hline 
    $T_1$     &  Targeted SINR threshold for UE$_1$  & 5  \\ \hline 
    $T_2$     &  Targeted SINR threshold for UE$_2$  & 5  \\ \hline 
    $T_3$     &  Targeted SINR threshold for UE$_3$  & 5  \\ \hline
    $\sigma^2$   &  AWGN power at BS receiver   & 1 \\ \hline
    $\beta\text{max}$ & Maximum A-RIS reflection coefficient & 1 \\ \hline
    $\mathbb{E} {\vert h_1\vert}$ &  Gain between UE$_1$ \& BS   & 5    \\ \hline 
    $\mathbb{E} {\vert h_2\vert}$ & Gain between UE$_2$ \&  BS  & 2 \\ \hline 
    $\mathbb{E} {\vert h_3 \vert}$ &  Gain between UE$_3$ \&  BS  & 1  \\ \hline 
    $\mathbb{E}{ \vert \fbf(n)\vert}$  &  Gain between A-RIS \&   BS   & 2  \\ \hline 
    $\mathbb{E}{\vert \gbf_1(n)\vert}$ & Gain between UE$_1$ \&  A-RIS & 1  \\ \hline 
    $\mathbb{E}{\vert \gbf_2(n)\vert}$ &  Gain between UE$_2$ \&  A-RIS   & 1   \\ \hline 
    $\mathbb{E}{\vert \gbf_3(n)\vert}$  & Gain between UE$_3$ \&  A-RIS   & 0.2 \\ \hline
    $\mathbb{E}{\vert \hbf_\text{j}(n)\vert}$  & Gain between jammer \&  BS & 1 \\ \hline
    $\mathbb{E}{\vert \Gbf_\text{j}(n,m)\vert}$ & Gain between jammer \& A-RIS & 0.2  \\ \hline 
    \end{tabular}
\end{table}

\newpage We choose to focus on a jammer with a rather large number of antennas $M$ and also an A-RIS with a large number of elements $N$. This is an interesting case in practice, since in mm-wave systems a rather large number of antennas/elements would be needed, and we want to show the capability of Algorithm~\ref{alg:PLS} to find a solution for such large optimization problems. This is in contrast to our initial work in \cite{b25} using the Matlab \emph{fmincon} tool, in which we could not find an optimal solution for a jammer larger than $M=16$ in combination with A-RIS larger than $N=32$.
Note that we assume a maximum A-RIS reflection coefficient $\beta_\text{max}=1$, despite our discussions in Section~\ref{sec:Absorb}. The reason is that in our considered evaluation scenarios, a value of $\beta_\text{max}<1$ can simply be included in the assumed path losses to the A-RIS.

\begin{figure}
    \centering
    \includegraphics[height=6cm]{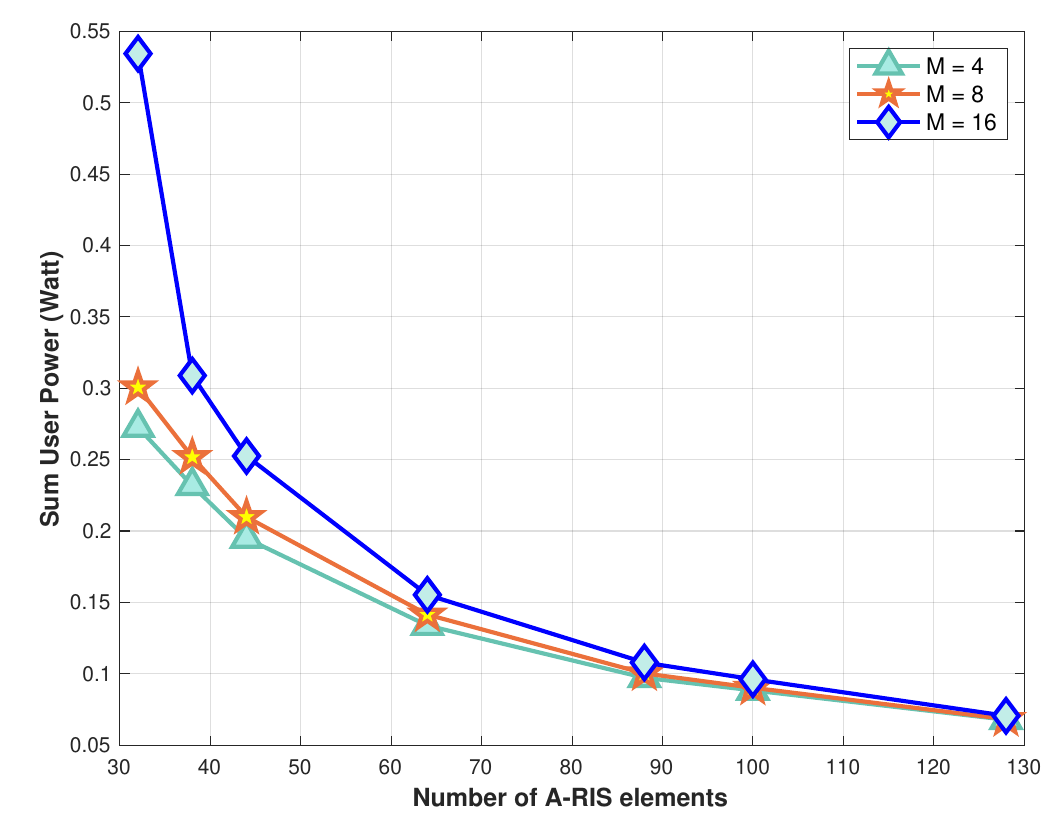}
    \caption{Required total transmit power to meet the user quality of service requirements in the presence of a jammer with $M=4, 8, 16$ antennas, as a function of the number of A-RIS elements $N=32, 38, 44, 64, 88, 100, 128$ in case of $K=2$ users.}
    \label{fig_psum_vs_N_K2}
\end{figure}
In Fig.~\ref{fig_psum_vs_N_K2} we show the required total user transmit power to meet the user QoS requirements in the presence of a jammer with $M=4, 8, 16$ antennas, as a function of the number of A-RIS elements $N=32, 38, 44, 64, 88, 100, 128$ for $K=2$ users. As seen, a large A-RIS is very beneficial to mitigate the jammer's power. The sum transmit power drops from 0.28 W (24.5 dBm) to 0.07 W (18.5 dBm) when increasing the number of A-RIS elements from $=4$ to $N=128$ in the case of a jammer with $M=4$ antenna elements. Doubling the number of jammer antennas to $M=8$ does not significantly change the required total transmit power. However, with $M=16$ there is a significant increase in the required sum user power for A-RIS with $32$ elements, increasing to 0.54 W (27.3 dBm), whereas for large A-RIS the increase in sum user power is still small.

\begin{figure}
    \centering
    \includegraphics[height=6cm]{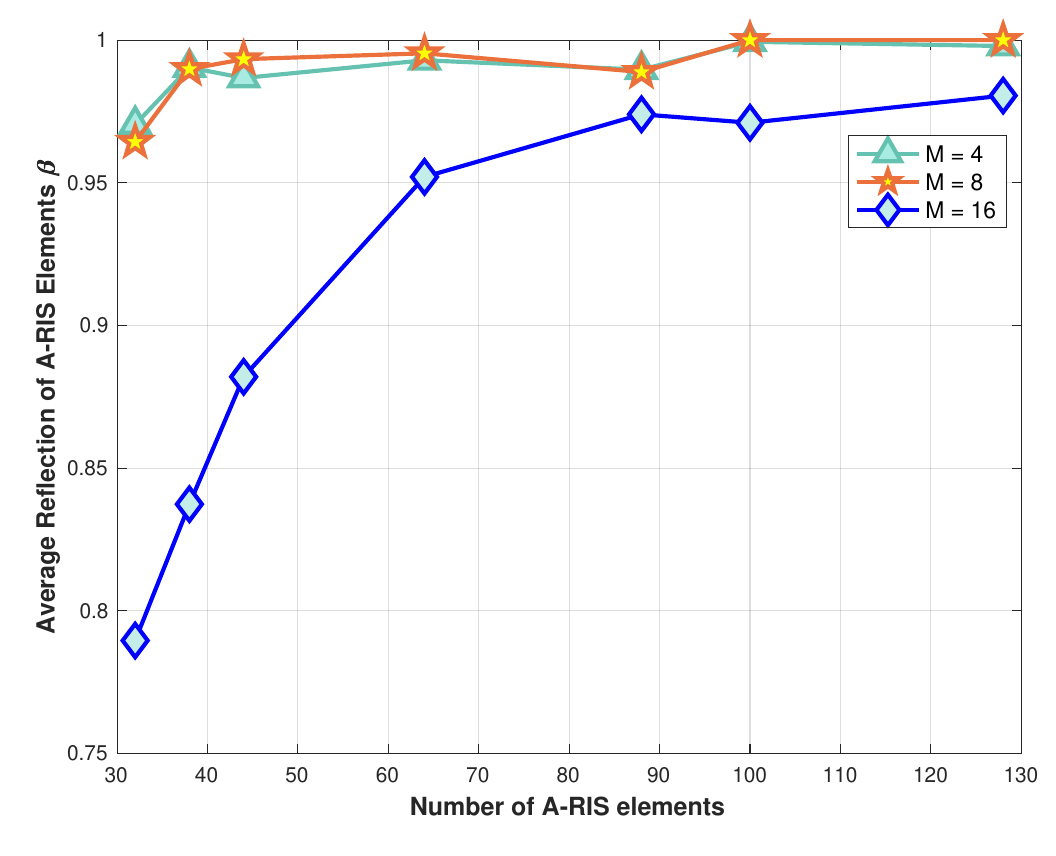}
    \caption{Resulting average reflection as a function of the number of A-RIS elements $N=32, 38, 44, 64, 88, 100, 128$ for $K=2$ users in the presence of a jammer with $M=4, 8, 16$ antennas.}
    \label{fig_meanbeta_vs_N_K2}
\end{figure}
As seen in Fig.~\ref{fig_meanbeta_vs_N_K2}, the reason for this is that absorption starts to be needed to mitigate the jammer: In the case of $N=32$ and $M=4, 8$ there is almost no absorption by the A-RIS with an average reflection close to $\bar \beta \approx 1$, whereas with $M=16$ the average reflection gain drops to $\bar \beta=0.8$ for $N=32$, and then gradually increases to $\bar \beta \approx 1$ as $N$ increases.

\begin{figure}
    \centering
    \includegraphics[height=6cm]{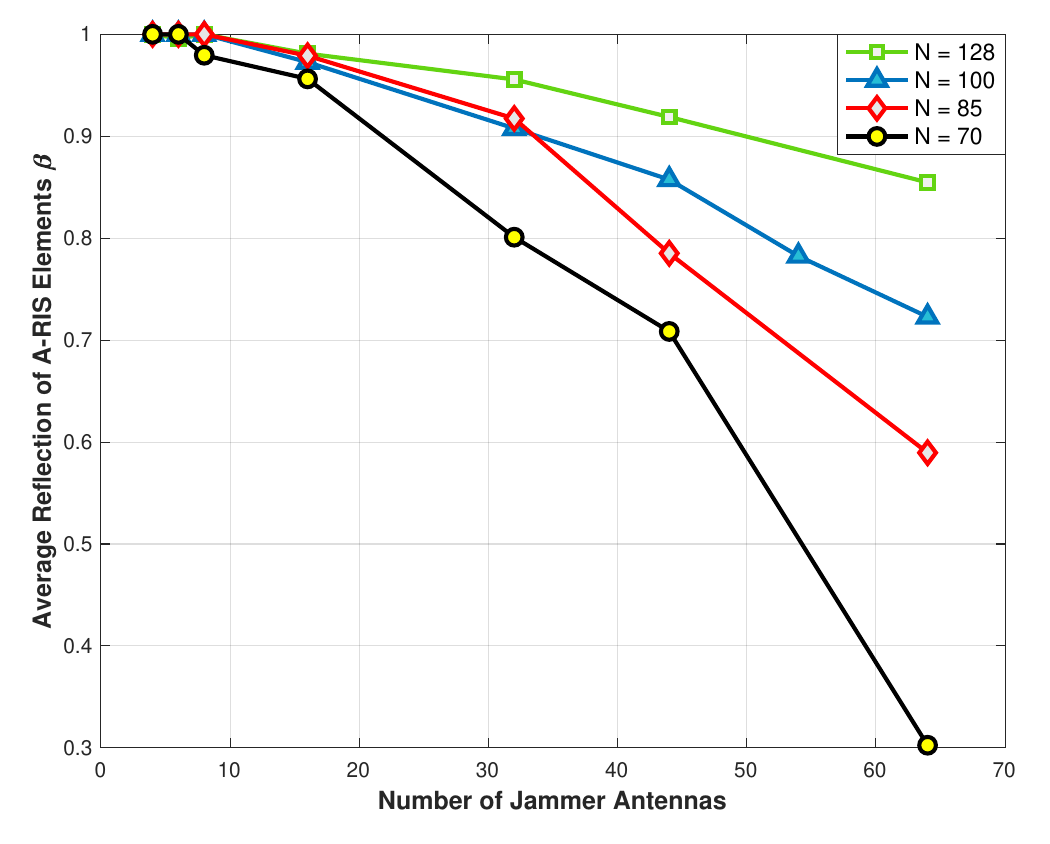}
    \caption{Resulting average reflection level for jammers $M=4,6,8,16,32,44,54,64$ antennas for $70, 85, 100, 128$ A-RIS elements in case of $K=2$ users.}
    \label{fig_meanbeta_vs_M_K2}
\end{figure}

In Fig.~\ref{fig_meanbeta_vs_M_K2}, we further investigate the absorption behavior of the A-RIS as a function of the number of jammer antennas $M$ and A-RIS elements $N$. Here we investigate larger $M$ and $N$ compared to Fig.~\ref{fig_meanbeta_vs_N_K2}, but the behavior of the A-RIS is the same. This also shows that Algorithm~\ref{alg:PLS} converges well also for large $M$ and $N$. In general, for the investigated deployment and propagation scenario here, it seems that with $N/M \geq \approx 5$, absorption is not needed, whereas for $N/M < \approx 5$, absorption is very useful. This can be explained by the required number of DoFs for the A-RIS to cancel the jammer signal.

\begin{figure}
    \centering
    \includegraphics[height=6cm]{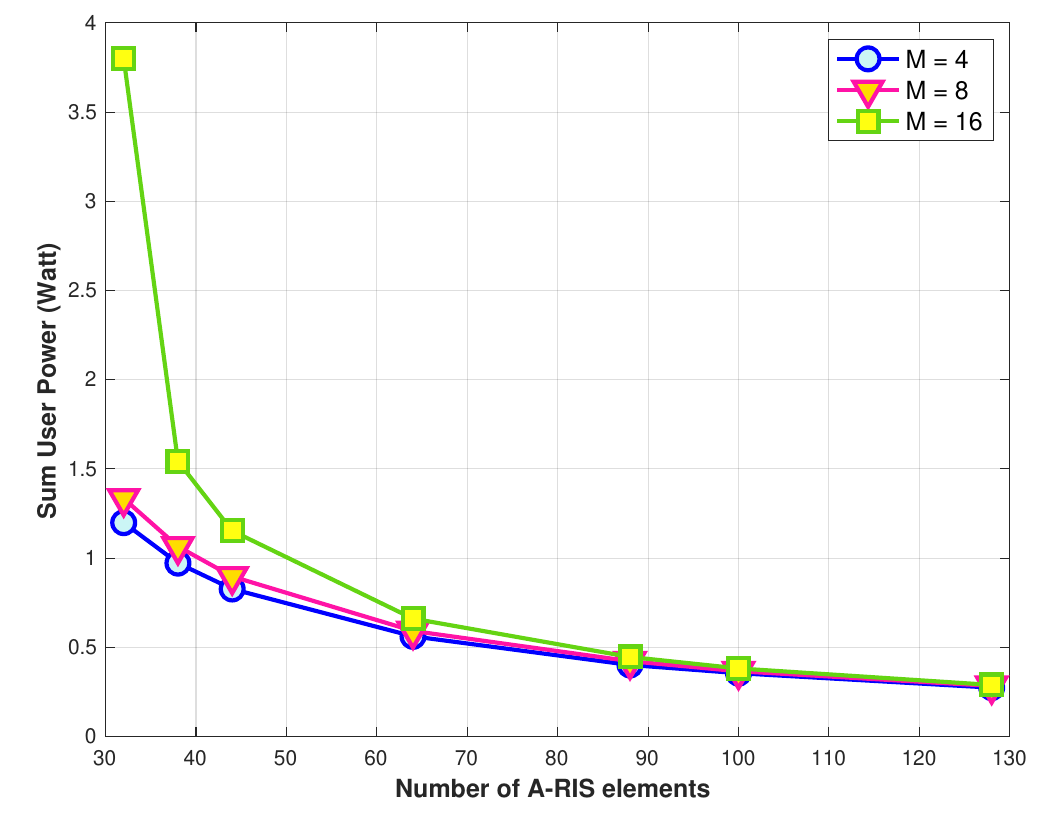}
    \caption{Required total transmit power to meet the user QoS requirements in the presence of a jammer with $M=4, 8, 16$ antennas, as a function of the number of A-RIS elements in case of $K=3$ users.}
    \label{fig_psum_vs_N_K3}
\end{figure}

\begin{figure}
    \centering
    \includegraphics[height=6cm]{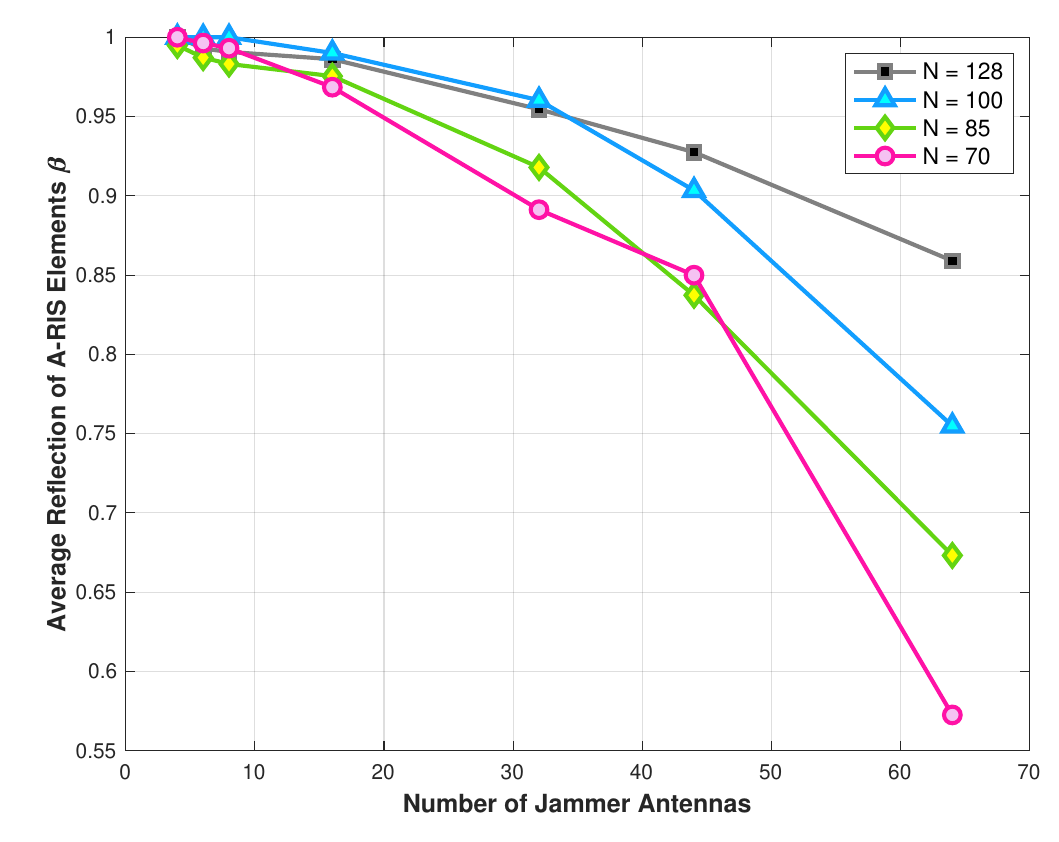}
    \caption{Resulting average reflection of A-RIS for jammers with $M=4, 6, 8, 16, 32, 44, 64$ antennas for $70, 85, 100, 128$ A-RIS elements in case of $K=3$ users.}
    \label{fig_meanbeta_vs_M_K3}
\end{figure}

In Fig.~\ref{fig_psum_vs_N_K3} and Fig.~\ref{fig_meanbeta_vs_M_K3}, we investigate the performance and behavior of the A-RIS in the case of $K=3$ users. Algorithm~\ref{alg:PLS} also converges very well in this case, and the overall behavior is the same as in the $K=2$ users case. One difference comparing Fig.~\ref{fig_psum_vs_N_K3} with Fig.~\ref{fig_psum_vs_N_K2} is that the total required transmit power by the users is larger for $K=3$ compared to $K=2$, due to the residual NOMA multi-user interference for the users $k<K$ that is not canceled at the BS. Comparing Fig.~\ref{fig_meanbeta_vs_M_K2} and Fig.~\ref{fig_meanbeta_vs_M_K3} shows that the A-RIS absorption \textit{decreases} for the larger number of users with increased multiuser interference (MUI). This is interesting since absorption \textit{increases} when the jammer interference increases. The difference can be explained by the fact that the A-RIS is attempting to entirely cancel the jammer, while combating the MUI still requires the individual user signals to be received with a strong SNR.

\begin{table}
    \centering
     \caption{Comparison of required total power in the presence of a jammer with $M=4, 8, 16$ antennas for $K=2, 3, 4$ users, with and without A-RIS ($N=32$)}
    \begin{tabular}{ccccc}
        \toprule
        \textbf{Number of $M$} & \textbf{Number of UE} & \textbf{Without A-RIS} & \textbf{With A-RIS} \\ \hline \hline
        {$M=4$}  & 2 Users  & $0.27 \times 10^3$ & $0.27$  \\ 
                 & 3 Users  & $0.51 \times 10^4$ & $1.19$  \\ 
                 & 4 Users  & $1.65 \times 10^4$ & $10.68$ \\ \midrule
        {$M=8$}  & 2 Users  & $0.59 \times 10^3$ & $0.30$  \\ 
                 & 3 Users  & $0.89 \times 10^4$ & $1.33$  \\ 
                 & 4 Users  & $1.35 \times 10^4$ & $11.90$ \\ \midrule
        {$M=16$} & 2 Users  & $1.07 \times 10^3$ & $0.54$  \\ 
                 & 3 Users  & $1.90 \times 10^4$ & $3.80$  \\ 
                 & 4 Users  & $2.72 \times 10^4$ & $36.53$ \\ \bottomrule
    \end{tabular}
      \label{table2}
\end{table}

In Table \ref{table2} we present a comparison of the total required transmit power in the presence of a jammer with different numbers of antennas $M=4, 8, 16$ for $K=2$, $K=3$ and $K=4$ users, with and without the assistance of an A-RIS. As shown, there is a very large gain with the A-RIS, up to more than a $28$ dB gain in the required total transmit power. The results also clearly show that as the number of jammer antennas $M$ increases, the total power consumption increases more without an A-RIS than with an A-RIS. When increasing the number of users $K$ from $K=3$ to $K=4$, there is a larger increase in the required total transmit power than going from $K=2$ to $K=3$ users, indicating that using the A-RIS degrees of freedom to simultaneously mitigate the jammer and control the multi-user interference is getting more challenging. However, the gain compared to the case with no A-RIS is still substantial.

\subsection{Performance for Various A-RIS Locations}
\label{sec:VariousARISLocations}
In order to understand the optimal behavior of the A-RIS under various propagation conditions, we analyze the performance of a range of A-RIS locations, as illustrated in Fig.~\ref{MovingRIS}. By changing the locations of the A-RIS, the path loss for all the channels to/from the A-RIS changes. We initially locate the A-RIS closer to the BS than any of the $K$ users, and also closer than the jammer. Then we gradually move the A-RIS further away from the BS towards the jammer and the users, until the distance to the BS becomes larger compared to the distance to the BS for the jammer and any of the $K$ users.
\textbf{\begin{figure}
    \centering
    \includegraphics[height=6cm]{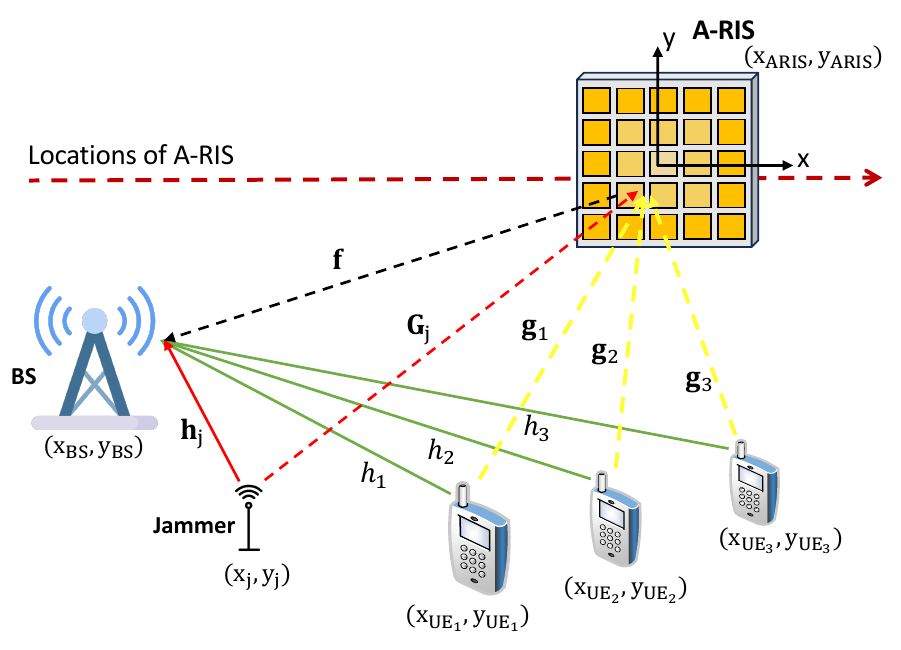}
    \caption{Illustration of an A-RIS deployment with gradually larger distance to the BS compared to the jammer and the $K$ users.}
    \label{MovingRIS}
\end{figure}
}
\begin{table}
    \caption{Summary of Deployment Parameters with Moving A-RIS. \label{table3} }
    \centering
    \tabcolsep=0.10cm
    \begin{tabular}{l|l|l}
    \bf{Variable}    &  \bf{Description}  & \bf{Value} \\ \hline \hline
    $K_0$ & Path loss factor & 3e5 \\ \hline
    $\alpha_1$  &  Path loss exponent for $h_1$, $h_2$, $h_3$  & 3\\ \hline
    $\alpha_2$  &  Path loss exponent for $\hbf_\text{j},\Gbf_\text{j}, \fbf, \gbf_1$, $\gbf_2$, $\gbf_3$ & 2\\ \hline
    $x_\text{BS}, y_\text{BS}$ & Location of BS (m) & 0, 0 \\ \hline
    $x_\text{ARIS}, y_\text{ARIS}$ &  Locations of A-RIS (m) & 10-200, 65 \\ \hline
    $x_\text{j},  y_\text{j}$ & Location of jammer (m) & 50, -80 \\ \hline 
    $x_{\text{UE}_1}, y_{\text{UE}_1}$ & Location of UE$_1$ (m) & 30, -15 \\ \hline
    $x_{\text{UE}_2}, y_{\text{UE}_2}$ & Location of UE$_2$ (m) & 50, -30 \\ \hline
    $x_{\text{UE}_3}, y_{\text{UE}_3}$  & Location of UE$_3$ (m) & 80, -45 \\ \hline
\end{tabular}
\end{table}

We adopt a two-dimensional deployment model described by the ($x$, $y$) positions of the nodes, as defined in Fig.~\ref{MovingRIS}. We then position the A-RIS at various locations along the $x$-axis, with the BS at the origin, and the jammer and users at the fixed locations given in Table~\ref{table3}. The related path losses are then calculated for all the channel components. 

We use the path loss model $P_r/P_t=K_0/d^\alpha$, in which we assume path loss exponent $\alpha_1=3$ for all UE-BS channels, and free space path loss, $\alpha_2=2$, for all other channels. These path loss exponents can be motivated by assuming the A-RIS is positioned in a location with favorable propagation conditions, and assuming the jammer also would strive for favorable propagation conditions. For numerical stability in the optimization, we normalize the noise at the BS as $\sigma^2=1$ and scale the path loss factor $K_0$ correspondingly. All other non-channel related parameters are the same as in Table~\ref{table1}.

\begin{figure}
    \centering
    \includegraphics[height=6cm]{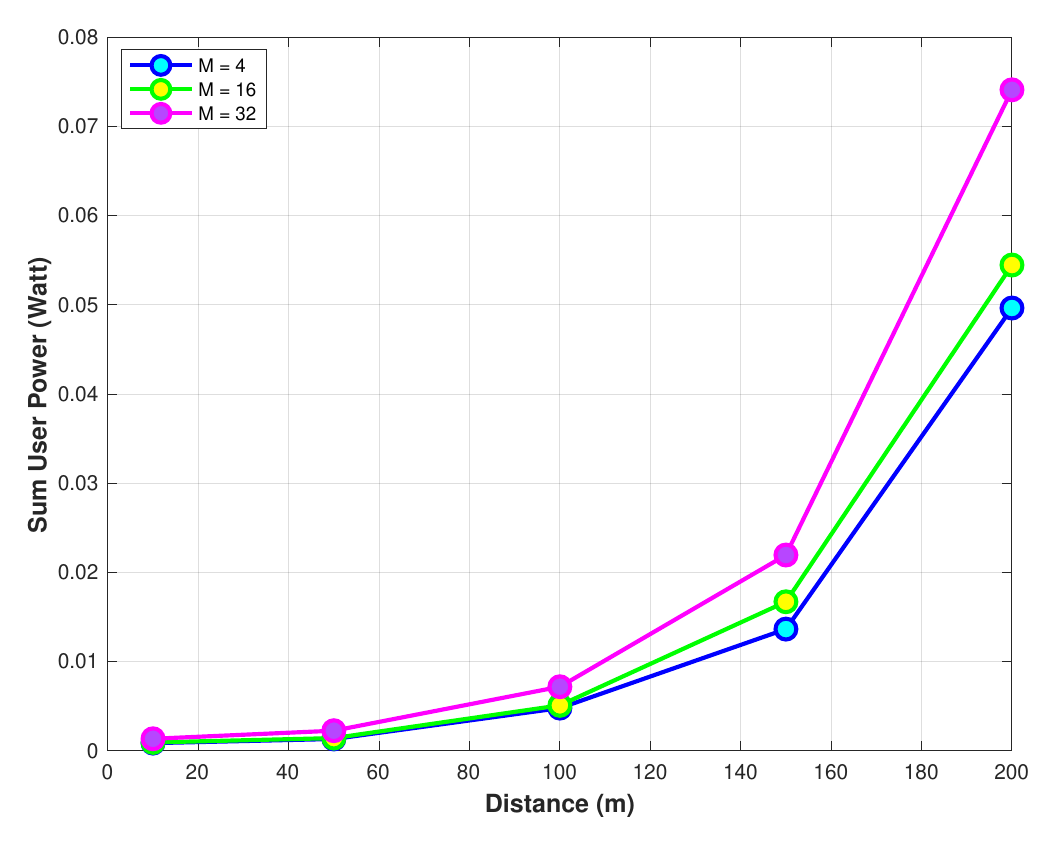}
    \caption{Required total transmit power by moving A-RIS with $N = 64$ elements away from the BS with jammer having $M=4, 16, 32$ antennas, as a function of distance in case of $K=3$ users. Distance refers to the coordinate along the $x$-axis.}
    \label{fig_psum_vs_dist_K3}
\end{figure}

\begin{figure}
    \centering
    \includegraphics[height=6cm]{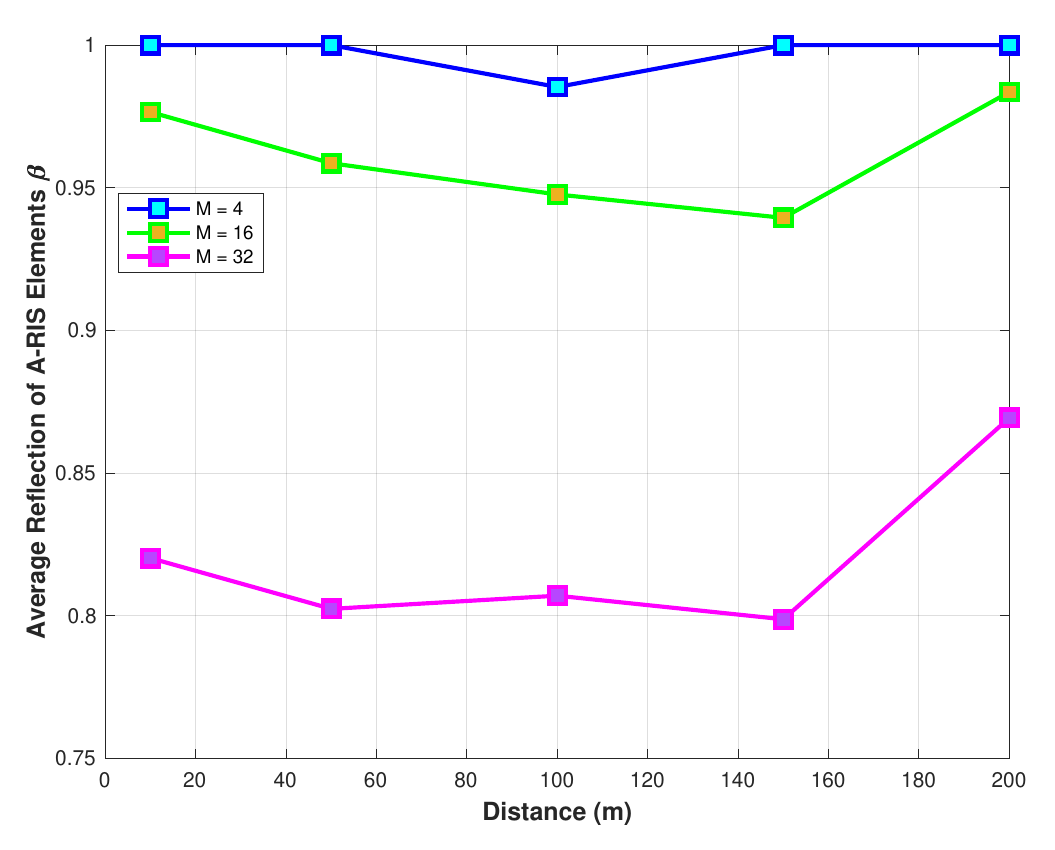}
    \caption{Average amplitude of the A-RIS reflection coefficients as the A-RIS moves away from the BS for a case with $K=3$ users, $N=64$ A-RIS elements, $M=4, 16, 32$ jammer antennas. Distance refers to the coordinate along the $x$-axis.}
    \label{fig_absorb_vs_dist_K3}
\end{figure}

In Fig.~\ref{fig_psum_vs_dist_K3}, we show the resulting required total transmit power as a function of the distance of the A-RIS from the BS in the presence of a jammer. We assume $N=64$ A-RIS elements and a jammer with $M=4, 16, 32$ antennas. As can be seen, the required total transmit power monotonically increases when the A-RIS is moving away from the BS.

Fig.~\ref{fig_absorb_vs_dist_K3} plots the average amplitude of the optimal A-RIS reflection coefficients as a function of the distance of the A-RIS from the BS. As in Fig.~\ref{fig_psum_vs_dist_K3}, we assume $N=64$ A-RIS elements and the three curves are for a jammer with $M=4, 16, 32$ antennas, respectively. As can be seen, the absorption is larger when the A-RIS is closer to the jammer and decreases as the A-RIS moves further away from the jammer and the users. We also see the effect discussed in the previous section, in which the absorption by the A-RIS increases with an increase in the dimension of the jammer array $M$.  

\subsection{Relevance of the Results for Downlink Scenario}
\label{sec:DownlinkScenario}

\textbf{\begin{figure}
    \centering
    \includegraphics[height =6.5cm]{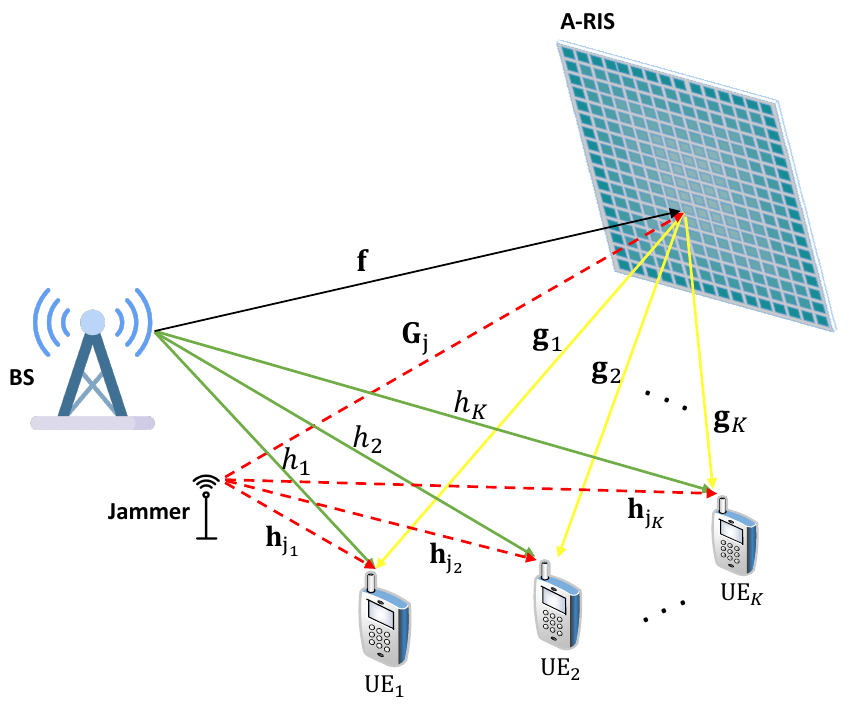}
    \caption{Illustration of an A-RIS-aided downlink NOMA system with $K$ users and a jammer.}
    \label{Downlink}
\end{figure}
}

Here we briefly explore a corresponding downlink scenario, as illustrated in Fig.~\ref{Downlink}, in order to investigate if there is an \textit{uplink-downlink} duality. In general, the downlink problem is quite different, since the jammer has more than one potential target, more precisely, all the $K$ users. Thus, to analyze the downlink problem, we would need to define the objective of the jammer:
\begin{itemize}
    \item Case 1: Select only one victim user
    \item Case 2: Attack the users with some kind of priority
    \item Case 3: Attack all the users with equal importance
\end{itemize}

In Case 1, the jammer will optimize its signal, focusing on attacking only a selected victim receiver. This case would be similar to the uplink case, in which the jammer focuses on optimizing its attack only on the BS. However, a key difference is that in our uplink model, there is only one receiver, the BS, whereas in the downlink, there are still $K-1$ other active receivers besides the one under attack that would still experience the jammer interference. In addition, since in general the path losses $\|\hbf_{\text{j}, k}\|$ are different between the jammer and the $K$ users, there is no straightforward way to map the power minimization problem in the downlink to Algorithm~\ref{alg:PLS} due to point-to-multipoint channels from the jammer to the users, resulting in a user index $k$-dependent jammer power $\sigma_\text{j}^2(k_v, k)$, where  $k_v$ denotes the victim user under jamming attack.

In Case 2, one could define the jamming signal as a weighted sum of the jamming signals targeting each user. In that case, it is straightforward to show that the jamming power at user $k$ would be the same for all victims $k_v$, but would still depend on $k$, i.e. $\sigma_\text{j}^2(k)$.
 
In Case 3, one could define the objective of the jammer to maximize $\sigma_\text{j}^2(k_v, k)=\sigma_{j}^2$, i.e., equal received jammer power for all users, independent of $k$. For propagation conditions in which there would be a feasible solution for the jammer to meet this objective, we would end up with an optimization problem that could potentially be solved using Algorithm~\ref{alg:PLS}.

\section{Conclusion} \label{sec:Conclusion}

In this paper, we have considered an uplink power-domain NOMA system assisted by an A-RIS in the presence of an intelligent jammer. An A-RIS can adaptively adjust both the amplitude and phase shift of its elements, providing more control of the surface, particularly for interference mitigation scenarios. A realistic model for constructing an A-RIS with independently tunable amplitude and phase was presented, together with important hardware characteristics that should be considered in realistic implementations.

An optimization problem was formulated whose goal is to minimize the total transmit power of the users under constraints on the SINR that the users achieve at the base station. The optimization is non-convex and of high dimension, so we reformulate it and employ an iterative approach that alternates between using linear programming to estimate the user powers and the Dinkelbach algorithm to determine the A-RIS coefficients.

Our simulation results demonstrate that A-RIS elements provide dramatic reductions in required total transmit power due to their ability to enhance the signals of interest and cancel the effects of the interference. We also demonstrated that if the number of elements in the intelligent surface is large, the A-RIS can provide additional degrees of freedom that enable increased signal enhancement and interference mitigation. We showed that the influence of the absorption is most important when the ratio of A-RIS elements to jammer antennas is not too large, i.e., when the extra degrees of freedom offered by the variable element amplitudes become crucial. Finally, we briefly discussed the relevance of our results for a corresponding downlink scenario.

\section*{Acknowledgment}
This work was supported by the project SEMANTIC, funded by EU's Horizon 2020 research and innovation programme under the Marie Skłodowska-Curie grant agreement No 861165, as well as by the {U.S.} National Science Foundation under grants CNS-2107182 and ECCS-2030029.

\ifCLASSOPTIONcaptionsoff
  \newpage
\fi

\begin{IEEEbiography}[{\includegraphics[width=1.1in,height=1.25in,clip,keepaspectratio]{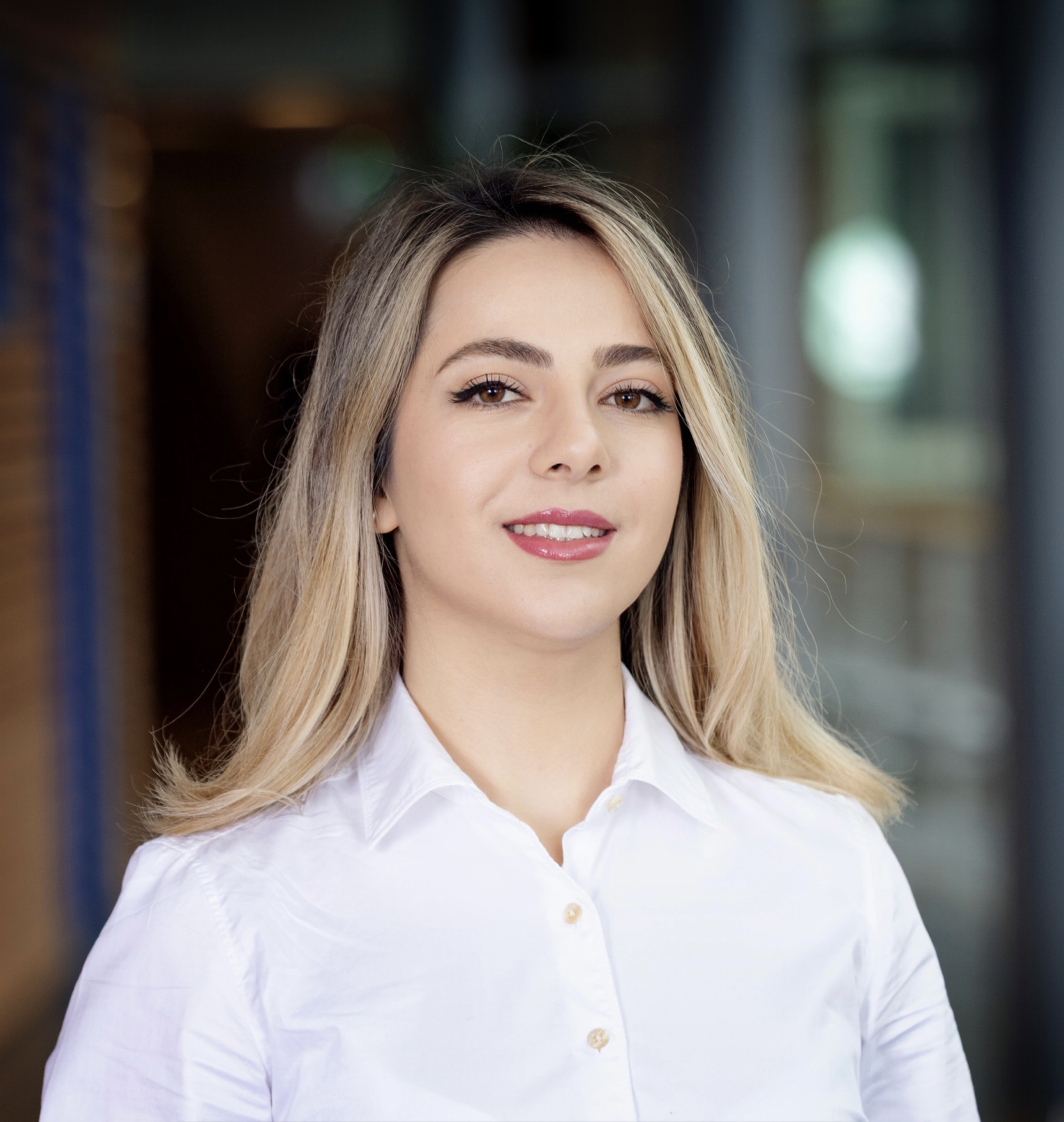} 
}]
{Azadeh Tabeshnezhad} (Student Member, IEEE) 
received her M.Sc. degree in Communication Systems from Science and Research Branch, Tehran, Iran, in 2018. She is currently pursuing a Ph.D. degree at Chalmers University of Technology, Sweden. She was a visiting researcher at UCI in 2022, 2024, and 2025.
Her research interests include NOMA, (Non)-convex optimization, reconfigurable intelligent surfaces, and ISAC. 
\end{IEEEbiography}

\begin{IEEEbiography}[{\includegraphics[width=1in,height=1.25in, clip,keepaspectratio]{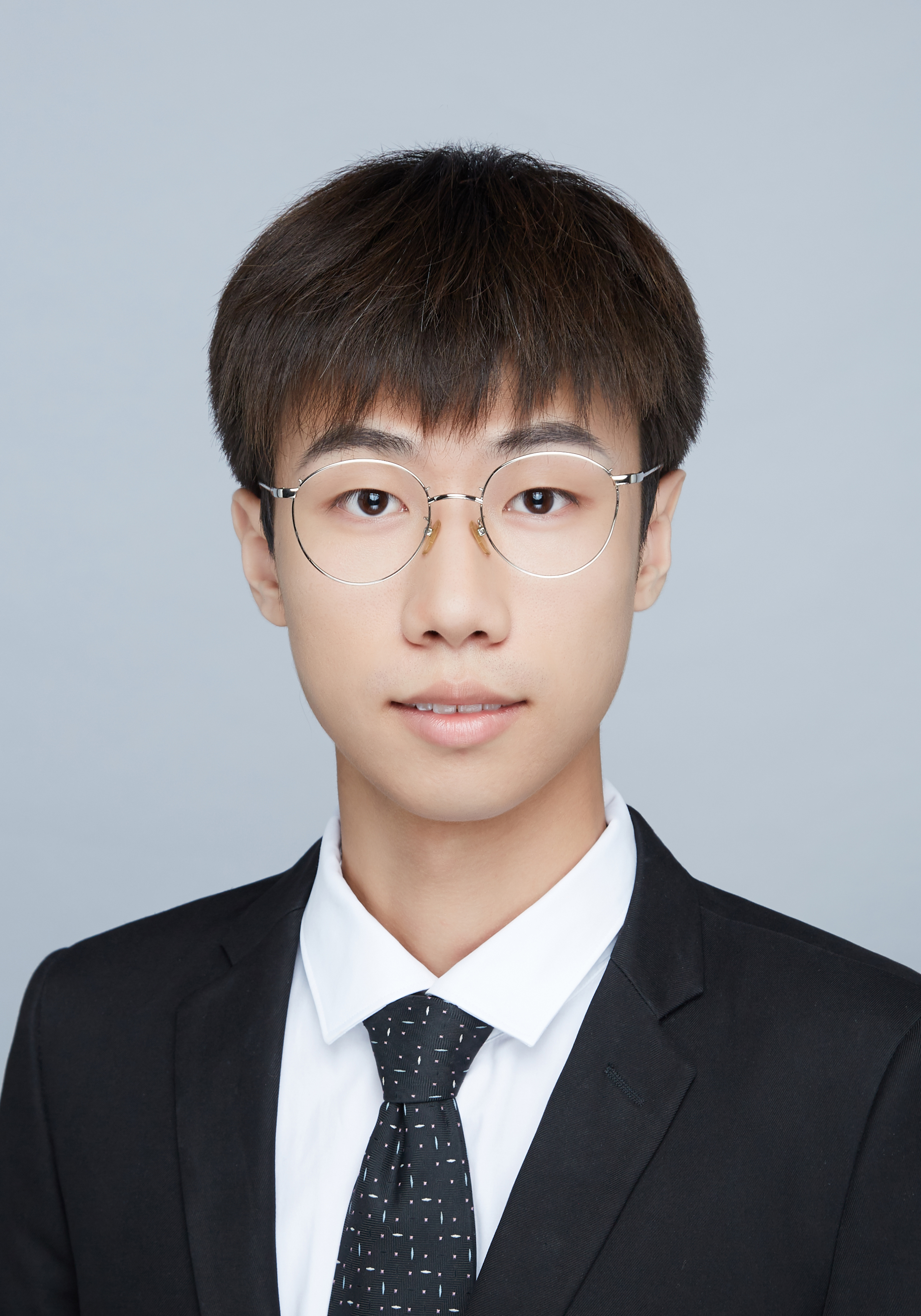}}]{Yuqing Zhu} (Graduate Student Member, IEEE) received the B.S. degree in electronic
 information engineering and the M.S. degree in electronics science and technology from Beijing
 Institute of Technology, Beijing, China, in 2019, and
 2022, respectively.
He is currently pursuing a Ph.D. degree with the Antenna Systems Group, Department of Electrical Engineering, Chalmers University of Technology, Gothenburg, Sweden. His current research interests include millimeter-wave antennas and arrays, phased array antennas, 5G/6G mobile terminal antennas, reconfigurable intelligent surfaces, and over-the-air testing techniques.
Mr. Zhu was a recipient of the Oral Best Paper Award at ICEICT 2020, the Best Student Paper Award at ICMMT 2022, and the Student Travel Grant at AP-S/URSI 2024 from the IEEE Antennas and Propagation Society. He serves as a reviewer for \textsc{IEEE Transactions on Antennas and Propagation}.
\end{IEEEbiography}

\begin{IEEEbiography}[{\includegraphics[width=1in,height=1.25in,clip,keepaspectratio]{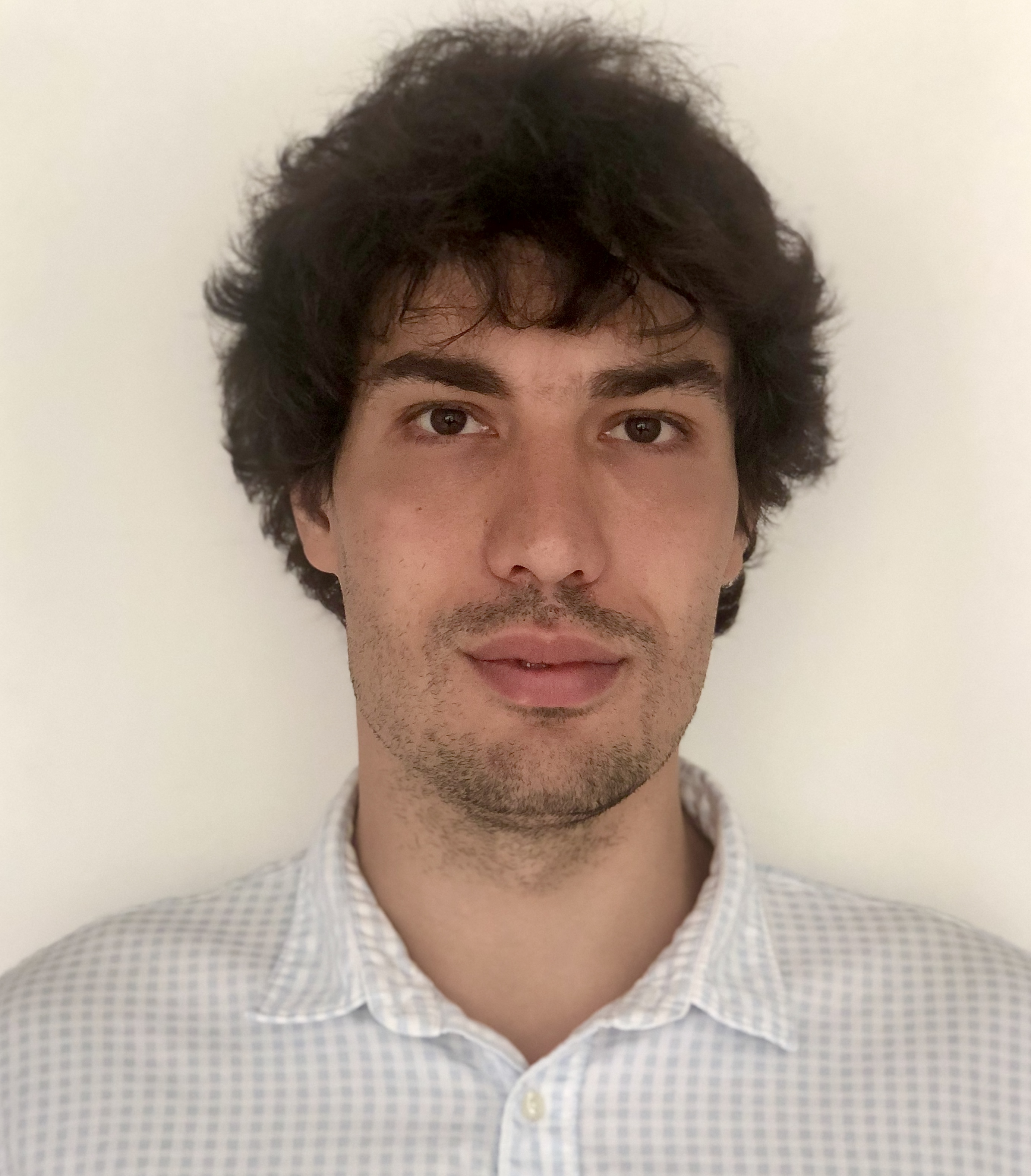}}]{Artem R. Vilenskiy}(Senior Member, IEEE) 
    received the specialist degree (\emph{summa cum laude}) in radio engineering and the Ph.D. degree in antennas and microwave devices from Bauman Moscow State Technical University, Moscow, Russia, in 2011 and 2014, respectively.
    From 2011 to 2019, he was with Samsung Research Institute Russia, Moscow, Russia, where he had the role of researcher, expert engineer, and project leader in the Electromagnetics Group and RF Sensor Part. From 2015 to 2019, he was an Associate Professor at the Radio-Electronic Systems and Devices Department of Bauman Moscow State Technical University. Since 2019, he has been a Researcher with Antenna Systems Group, Chalmers University of Technology, Gothenburg, Sweden. His current research interests include array antennas, periodic structures, integrated circuit design, controllable high-frequency materials and devices, as well as computational electromagnetics.
\end{IEEEbiography}

\begin{IEEEbiography}[{\includegraphics[width=1.1in,height=1.25in,clip,keepaspectratio]{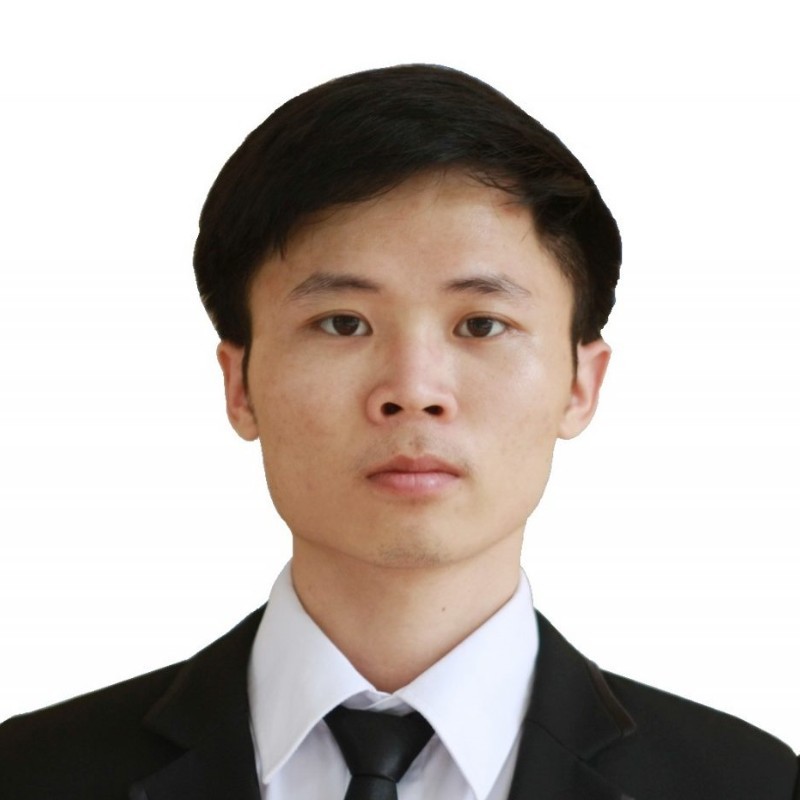}
}]
{Ly V. Nguyen}(Senior Member, IEEE)
received his M.Sc. degree in wireless communications from CentraleSupélec, Paris-Saclay University, France, in 2016, and the Ph.D. degree in Computational Science from the University of California, Irvine (UCI) and San Diego State  University (SDSU), USA, in 2022. He was a postdoctoral researcher at Purdue University and the University of California, Irvine (UCI) in 2022-25. Currently, He is an Assistant Professor in the Department of Electrical Engineering and Computer Science (EECS) and the Institute for Information Sciences (I2S) at the University of Kansas (KU).
His research interests lie in the areas of wireless communications, signal processing, and machine learning.
\end{IEEEbiography}

\begin{IEEEbiography}
[{\includegraphics[width=1.1in,height=1.25in,clip,keepaspectratio]{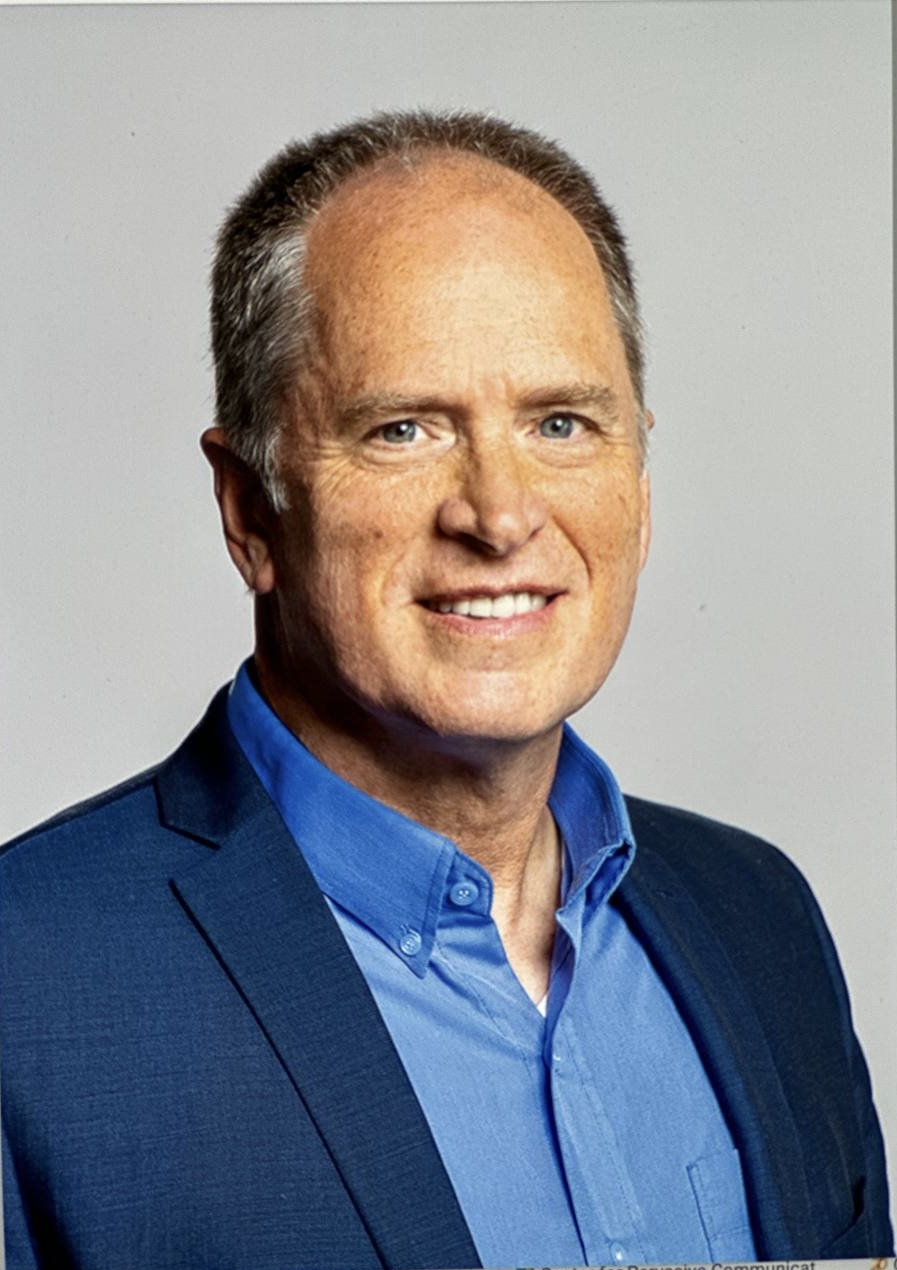} 
}]
{A. Lee Swindlehurst} (Fellow, IEEE)
received the B.S. (1985) and M.S. (1986) degrees in Electrical Engineering from Brigham Young University (BYU), and the PhD (1991) degree in Electrical Engineering from Stanford University. He was with the Department of Electrical and Computer Engineering at BYU from 1990-2007, where he served as Department Chair from 2003-06.  During 1996-97, he held a joint appointment as a visiting scholar at Uppsala University and the Royal Institute of Technology in Sweden. From 2006-07, he was on leave working as Vice President of Research for ArrayComm LLC in San Jose, California. Since 2007, he has been with the Electrical Engineering and Computer Science (EECS) Department at the University of California, Irvine, where he is a Distinguished Professor and currently serving as Department Chair. Dr. Swindlehurst is a Fellow of the IEEE, during 2014-17, he was also a Hans Fischer Senior Fellow in the Institute for Advanced Studies at the Technical University of Munich, and in 2016, he was elected as a Foreign Member of the Royal Swedish Academy of Engineering Sciences (IVA). He received the 2000 IEEE W. R. G. Baker Prize Paper Award, the 2006 IEEE Communications Society Stephen O. Rice Prize in the Field of Communication Theory, 2006, 2010 and 2021 IEEE Signal Processing Society’s Best Paper Awards, the 2017 IEEE Signal Processing Society Donald G. Fink Overview Paper Award, a Best Paper award at the 2020 and 2024 IEEE International Conferences on Communications, the 2022 Claude Shannon-Harry Nyquist Technical Achievement Award from the IEEE Signal Processing Society, and the 2024 Fred W. Ellersick Prize from the IEEE Communications Society. His research focuses on array signal processing for radar, wireless communications, and biomedical applications.
\end{IEEEbiography}

\begin{IEEEbiography}[{\includegraphics[width=1.1in,height=1.25in,clip,keepaspectratio]{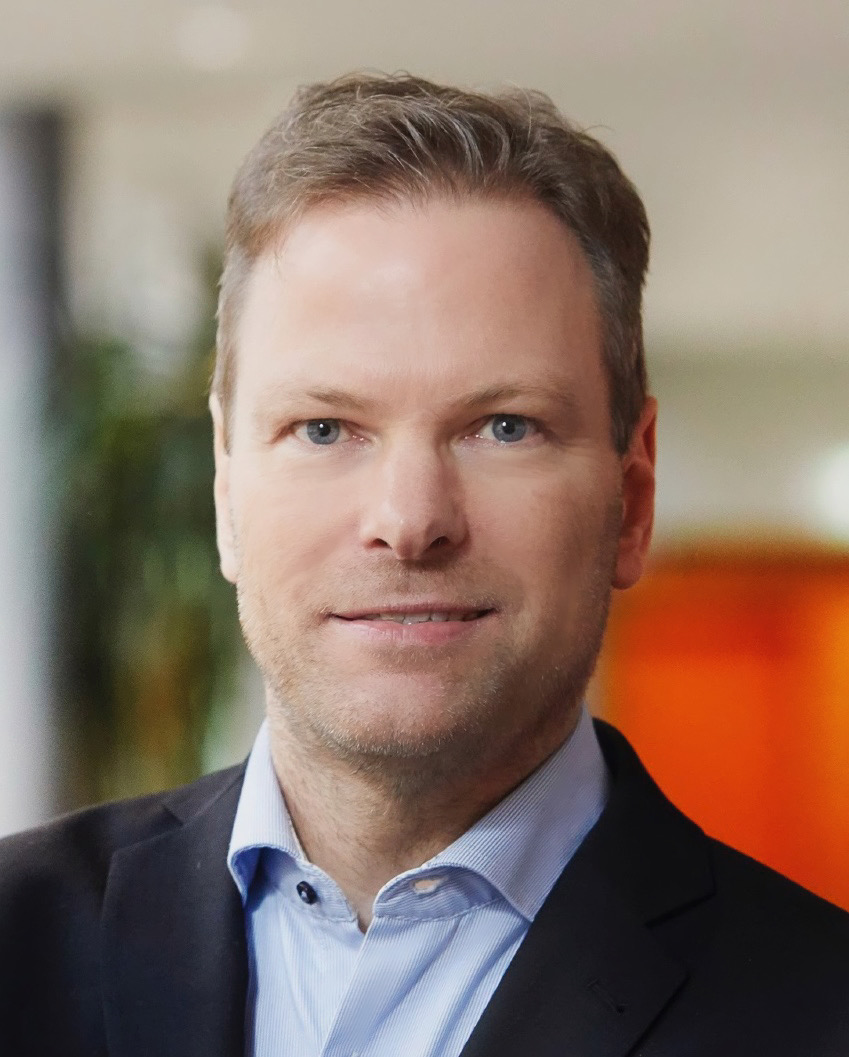} 
}]
{Tommy Svensson}(Senior, IEEE) is a Full Professor in Communication Systems at Chalmers University of Technology, leading the Wireless Systems research on air interface and wireless backhaul networking technologies. He received a Ph.D. in Information theory from Chalmers in 2003 and has worked at Ericsson AB with core, radio access, and microwave networks. He was active in the European WINNER I/II/+ and ARTIST4G projects with important contributions to the 3GPP LTE standards, the EU METIS and mmMAGIC and 5GCar projects towards 5G and the EU Hexa-X \& Hexa-X-II, RISE-6G, SEMANTIC, and newly started ROBUST-6G and ECO-eNET projects towards 6G, and in Chase/ChaseOn and newly started WiTECH antenna systems excellence centers at Chalmers targeting mm-wave and (sub)-THz solutions for access, backhaul/ fronthaul, and V2X scenarios.
His main research interests are in the design and analysis of mobile communication systems, physical layer algorithms, multiple access, resource allocation, cooperative/ context-aided/ secure communications, mm-wave/ sub-THz communications, C-V2X, JCAS, satellite networks, sustainable design, and end-to-end architecture.
He has co-authored 6 books, 115 journal papers, 156 conference papers, and more than 80 public EU projects' deliverables. He is a founding member/editor of the IEEE JSAC Series on Machine Learning in Communications and Networks and a board member of the Swedish Post and Telecom Authority (PTS).
He has been Chairman of the awards winning IEEE Sweden Vehicular Technology/ Communications/ Information Theory Societies chapter, editor of IEEE Transactions on Wireless Communications, IEEE Wireless Communications Letters, Guest editor of top journals, organized tutorials and workshops at top IEEE conferences, led local organizer of EuCNC \& 6G Summit 2023, and coordinator of the Communication Engineering Master’s Program at Chalmers.
\end{IEEEbiography}
\vfill
\end{document}